\documentclass[twocolumn,prd,nofootinbib]{revtex4}
\usepackage{graphicx}
\usepackage{dcolumn}
\usepackage{amsfonts}
\usepackage{amssymb}
\usepackage{amsmath}

\def\be{\begin{equation}}
\def\ee{\end{equation}}
\def\ba{\begin{eqnarray}}
\def\ea{\end{eqnarray}}

\begin{document}

\title{Spherically symmetric, static black holes with scalar hair, and naked
singularities in nonminimally coupled k-essence}
\author{Cec\'{\i}lia Nagy}
\affiliation{Institute of Physics, University of Szeged, D\'{o}m t\'{e}r 9, 6720 Szeged,
Hungary}
\author{Zolt\'{a}n Keresztes}
\affiliation{Institute of Physics, University of Szeged, D\'{o}m t\'{e}r 9, 6720 Szeged,
Hungary}
\author{L\'{a}szl\'{o} \'{A}. Gergely}
\affiliation{Institute of Physics, University of Szeged, D\'{o}m t\'{e}r 9, 6720 Szeged,
Hungary}
\date{\today }

\begin{abstract}
We apply a recently developed 2+1+1 decomposition of spacetime, based on a
nonorthogonal double foliation for the study of spherically symmetric,
static black hole solutions of Horndeski scalar-tensor theory. Our
discussion proceeds in an effective field theory (EFT) of modified gravity
approach, with the action depending on metric and embedding scalars adapted
to the nonorthogonal 2+1+1 decomposition. We prove that the most generic
class of Horndeski Lagrangians compatible with observations can be expressed
in this EFT form. By studying the first order perturbation of the EFT action
we derive three equations of motion, which reduce to those derived earlier
in an orthogonal 2+1+1 decomposition, and a fourth equation for the metric
parameter $\mathcal{N}$ related to the nonorthogonality of the foliation.
For the Horndeski class of theories with vanishing $G_{3}$ and $G_{5}$, but
generic functions $G_{2}\left( \phi ,X\right) $ (k-essence) and $G_{4}\left(
\phi \right) $ (nonminimal coupling to the metric) we prove the unicity
theorem that no action beyond Einstein--Hilbert allows for the Schwarzschild
solution. Next we integrate the EFT field equations for the case with only
one independent metric function and we obtain new solutions characterized by
a parameter, which in the simplest cases has the interpretation of mass or
tidal charge, the cosmological constant and a third parameter. These
solutions represent naked singularities, black holes with scalar hair or
have the double horizon structure of the Schwarzschild--de Sitter spacetime.
Solutions with homogeneous Kantowski--Sachs type regions also emerge.
Finally, one of the solutions obtained for the function $G_{4}$ linear in
the curvature coordinate, in certain parameter range exhibits an intriguing
logarithmic singularity lying outside the horizon. The newly derived hairy
black hole solutions evade previously known unicity theorems by being
asymptotically nonflat, even in the absence of the cosmological constant.
\end{abstract}

\maketitle

\section{Introduction\label{intro}}

Extensions of general relativity are well-motivated by its inability to
match astrophysical and cosmological observations unless dark matter and
dark energy are introduced, both undetectable otherwise than
gravitationally; the need for an inflationary universe with at least one
scalar added to the gravitational sector; or the study of the infrared limit
of the yet to be developed quantum theory of gravity. The addition of a
scalar field $\phi $ to the metric tensor as a gravitational variable is
among the simplest possible modifications. In order to avoid Ostrogradski
instabilities, such theories should be of second differential order for both
the scalar and the tensor. This condition is satisfied by the Horndeski
class of theories \cite{Horndeski,Deffayet}. A more encompassing class of
admissible theories restrict to second order only the evolution of the
degrees of freedom \cite{GLPVprl,GLPV}. As the validity of general
relativity has been accurately confirmed on the Solar System scale, there is
need for a mechanism to switch on any of its modifications only at a larger
radius. The viability of such a Vainshtein mechanism restricts the Horndeski
class \cite{Vainshtein1,Vainshtein2,Vainshtein3}. The recent confirmation,
at least at LIGO frequencies, that gravitational waves propagate with the
speed of light \cite{Multimessenger} adds further restrictions \cite%
{GWc1,GWc2,GWc3,GWc4}, leaving as viable theories only the so-called
generalized kinetic gravity braiding \cite%
{KineticBraidingKT,KineticBraidingDPSV} subset of Horndeski theories. This
subset however is still fairly rich, especially concerning the dynamics of
the scalar.

Such models were extensively investigated in a cosmological context,
including the junction along spacelike hypersurfaces \cite{PadillaS,NK}.
Similar junction conditions hold for timelike hypersurfaces, while
techniques developed for null hypersurfaces in general relativity \cite%
{BI,Poisson} were applied \cite{RG} for the case of null junction
hypersurfaces of kinetic gravity braiding models.

Black hole solutions were also sought for in these type of scalar-tensor
theories. The simplest of them is the Brans--Dicke theory \cite{BD}, in
which the gravitational constant scales with $\phi ^{-1}$ (where $\phi $ is
a massless scalar field) and a kinetic term $X=\partial ^{a}\phi \partial
_{a}\phi $ couples trough $\omega \phi ^{-1}$ (with $\omega $ a parameter)
to the curvature part of the action. Penrose has suggested as early as 1968
that gravitational collapse (discussed in the Einstein frame) proceeds
similarly \cite{Penrose1968}; hence, Brans--Dicke black holes are identical
to general relativistic ones \cite{Penrose1970} and his no-hair conjecture
has been supported in the large $\omega $ expansion scheme \cite{Thorne}.
Hawking had proven in 1972 that stationary black hole solutions of the
Brans--Dicke theory exist only for a constant scalar; hence, they are also
solutions of the Einstein field equations \cite{Hawking}. Key in the
derivation is the existence of a horizon, hence this result does not forbid
stationary stellar solutions different from their general relativistic
counterparts. In fact, Brans has found in 1962 all spherically symmetric,
static vacuum solutions of the Brans--Dicke theory \cite{BransPhD,Brans}.
However, none of those with $\phi \neq $const. possess event horizons;
instead they have naked singularities \cite{Thorne}. Solutions were also
found in the stationary, axisymmetric case as generalizations of the
Kerr--Newman family of solutions to the Brans--Dicke--Maxwell scenario \cite%
{TN,NT,SR}, and also solutions corresponding to the observer's two modes of
description of the static electromagnetic field: as axially symmetric
magnetic field or axially symmetric radial electric field \cite{SR2}. The
Kerr--Newman type solution was further analyzed \cite{Kim}. Interestingly,
when switching off both the rotation and the electric charge, it fails to
become spherically symmetric, a quadrupolar deformation (disappearing for $%
\omega \rightarrow \infty $, the general relativistic limit) being induced
by the scalar field, which itself exhibits a quadrupolar deformation from
spherical symmetry. The analysis has shown that the curvature invariants $R$%
, $R_{ab}R^{ab}$ and the Kretschmann scalar $R_{abcd}R^{abcd}$ vanish at the
horizon candidate $\Delta =0$ when $\omega \in \left( -5/2,-3/2\right) $,
while they diverge otherwise, and concluded that the Kerr--Newman type
solutions of the Brans--Dicke--Maxwell theory in the above range represent
black holes with scalar hair. However the rest of the curvature invariants
(there are 17 in total, related algebraically by 8 syzygies; see Ref. \cite%
{curvinv}) were not checked. They also remained unchecked for a spherically
symmetric and static black hole candidate in Brans--Dicke theory proposed in
Ref. \cite{CampanelliLousto}, based on the regularity (vanishing) of the
Kretschmann scalar on the horizon. The collapse of collisionless matter to a
black hole in Brans--Dicke theory has been also investigated numerically 
\cite{Teukolsky}. Contrary to general relativity, in Brans--Dicke theory the
Oppenheimer--Snyder collapse leads to a dynamical black hole, in which the
condition $R_{ab}l^{a}l^{b}\geq 0$ (for all null vectors $l^{a}$) assumed in
the derivation by Hawking can be violated; nevertheless, the end result is
the same as in general relativity \cite{Teukolsky2}.

The no-hair theorem for asymptotically flat, static, spherically symmetric
black holes in Brans--Dicke theory has been extended to multicomponent
scalar field configurations \cite{Bekenstein}. Numerical evidence ruling out
spherical scalar hair of static four-dimensional black holes has been
presented for scalar fields satisfying the Positive Energy Theorem, with a
potential derivable from a superpotential motivated by supergravity \cite%
{Hertog}.

Brans--Dicke theory has been generalized to include a potential (allowing
for massive scalar fields) and a scalar field dependent coupling $\omega
\left( \phi \right) $ (lifting the constraint $\omega >40000$ established
for a constant $\omega \,$ from the frequency shift of radio photons to and
from the Cassini spacecraft as they passed near the Sun \cite{omegaSolar}).
Question comes whether in this class of scalar-tensor theories stationary
black holes different from general relativistic ones could exist.

This has been investigated by Sotiriou and Faraoni, who extended Hawking's
result to stationary black holes in this class of theories \cite{Sotiriou}.
When there is no potential, however $\omega =\omega \left( \phi \right) \neq
-3/2$ and does not diverge, Hawking's original proof still holds. Otherwise
they show, that for stationary and isolated black holes (asymptotically flat
and with a constant $\phi _{0}$ value for the scalar field at spatial
infinity, implying the vanishing of the potential at infinity through the
tensorial equations of motion), by imposing linear stability for the scalar
in the Einstein frame, the scalar field ought to be a constant. Thus the
stationary solutions of these generalized Brans--Dicke theories include only
the general relativistic black holes.

By employing a 1+1+2 decomposition based on kinematical quantities, in the
particular case of a Klein--Gordon scalar field $\omega \left( \phi \right)
=\phi /2$ with arbitrary potential and coupled nonminimally to the metric,
Ref. \cite{Dunsby1+1+2} confirmed, that the Schwarzschild solution implies a
constant scalar field.

\textbf{\ }Hawking's no hair theorem in the particular case of spherically
symmetric, static black holes was also generalized for a wide class of
Hordenski theories dubbed as Galileon, which are invariant under the shift ($%
\phi \rightarrow \phi +$const.) of the scalar field \cite{Galileon}, as the
radial component of the Noether current vanishes ($J^{r}=0$) implying in
general the vanishing of the radial\ derivative of the scalar field ($\phi
^{\prime }=0$); hence, a constant scalar. A notable exception to this proof
was presented in Ref. \cite{GB1}, where a nontrivial Galileon couples to the
Gauss--Bonnet invariant. Another exception arises when the Galileon couples
to the Einstein tensor $G_{ab}$\ (in the form $G^{ab}\partial _{a}\phi
\partial _{b}\phi $), with the metric either does (has primary hair) or does
not deviate from the general relativistic solutions in the presence of a
nontrivial scalar (representing secondary hair) \cite{DerivativeCoupling}.%
\footnote{%
In terms of the Horndeski coefficients $G_{i}(\phi ,X)$ ($i=2,..,5$), where $%
X=\partial ^{a}\phi \partial _{a}\phi $ is the kinetic term, the occurrence
of the $G^{ab}\partial _{a}\phi \partial _{b}\phi $ coupling requires $%
\partial G_{4}/\partial X\neq 0$, while for the linear coupling to the
Gauss--Bonnet invariant, $G_{5}$\ has to include a contribution proportional
to $\ln \left\vert X\right\vert $.} This includes the Schwarzschild black
hole unaffected by a time and radial dependent scalar, an example of a
stealth black hole.\footnote{%
Other examples of stealth black holes, with the same metric as in general
relativity, but nonconstant scalar field configurations were discussed in
Refs. \cite{SBH1,SBH3,SBH4,SBH5,SBH6,SBH7,SBH8,SBH9,SBH10,SBH11}. Both
solutions with primary hair or secondary hair only (stealth black holes)
were identified in beyond Hordenski theories \cite{BeyondHordenski}.} Other%
\textbf{\ }hairy black holes with derivative coupling (to the Einstein
tensor) were found in Refs. \cite%
{DerivC0,DerivC1,DerivC2,DerivC3,DerivC4,DerivC5,DerivC6,GBDerivC7}, while
for coupling to the Gauss--Bonnet invariant in Refs. \cite%
{GB2,GB3,GB4,GB5,GB6}. Neither of these exceptional cases are however
observationally preferred, as the\textbf{\ }generalized kinetic gravity
braiding subset has $G_{4}\left( \phi \right) $ and $G_{5}=0$.

Both black holes with primary hair and stealth solutions with nontrivial
scalar have been identified in the framework of generalized kinetic gravity
braiding subclass with nonvanishing $G_{3}$ \cite{cubic}. They evade the
no-hair theorem of Ref. \cite{Galileon} also due to avoiding $\phi ^{\prime
}=0$ while $J^{r}=0$ holds.

Early investigations of stellar solutions showed that certain scalar-tensor
theories could pass the weak-field tests; however, the predictions in strong
field would differ from the general relativistic ones, opening the
opportunity of testing them with double pulsar experiments \cite{DEF1,DEF2}.
Indeed, the scalar field is sourced by the trace of the energy-momentum
tensor of the neutron star and this spontaneous scalarization affects the
dynamics of the double pulsar. The existence and modifications of
spherically symmetric and static neutron stars in both $f(R)$\ gravity and
Brans-Dicke theory with scalar potential were investigated for two
particular neutron star equations of state \cite{KT}\textbf{.} The stability
of spherically symmetric solutions of the cubic covariant Galileon model
with a matter source has been studied in the test scalar field approximation 
\cite{BEF}, while the stability of relativistic stars composed of perfect
fluid minimally coupled in the Jordan frame has been analyzed in Ref. \cite%
{StabRelStars} for a subclass of Horndeski theories with linear dependence
on the kinetic term.

In discussing geometries with spherical symmetry, it is natural to single
out a radial direction. Temporal evolution selects another direction; hence,
a 2+1+1 decomposition of spacetime could be useful. Such a\ formalism was
developed in Refs. \cite{s+1+1a,s+1+1b} and explored in the context of
braneworlds. It was based on geometrical quantities characterizing the
embedding of the 2-surface: extrinsic curvatures, normal fundamental forms
and normal fundamental scalars constructed with both singled-out vectors.
Some of them were related to temporal and radial derivatives of the metric,
playing an important role in the Hamiltonian treatment. The formalism of
Refs. \cite{s+1+1a,s+1+1b} was based on an orthogonal double foliation. This
unnecessary restriction was lifted in our previous work, Ref. \cite{GKG},
allowing for the nonorthogonal double foliation of spacetime. Hence, a new
degree of freedom, a measure of the nonorthogonality, represented by the
metric function $\mathcal{N}$ emerged, reestablishing generic gauge
invariance. The price to pay was that the number of geometric embedding
variables is doubled (as there are two orthogonal bases adapted to each
hypersurface normal); nevertheless the two sets were expressible in terms of
each other.

In this paper we apply this 2+1+1 decomposition of spacetime, based on a
nonorthogonal double foliation for the study of spherically symmetric,
static black hole solutions of Horndeski scalar-tensor theory. Our
discussion adopts an effective field theory (EFT) of modified gravity
approach, with the action depending on a set of scalars characterizing the
embedding of the two nonorthogonal foliations. We summarize the
nonorthogonal 2+1+1 formalism in Sec. II. and give the relation between the
embedding scalars emerging in the two bases in Appendix A.

In Sec. III we rewrite the Horndeski Lagrangians $\emph{L}_{2-4}$ into the
EFT form, by employing half of the embedding variables, those adapted to the
scalar field gradient.\footnote{%
We do not deal here with the most complicated contribution $\emph{L}_{5}$,
as it makes the Vainshtein mechanism unfeasible.} We prove in Appendix B the
rightness of this choice, as the Lagrangians are much more complicated in
terms of the other half of variables.

In Sec. IV we present the method to perform the first order variations of
the EFT action, depending on both metric and embedding variables. We also
give both the intrinsic and embedding variables for a spherically symmetric,
static background.

Section V contains the derivation of the equations of motion, following the
similar procedure developed for the orthogonal double foliation formalism 
\cite{KGT}. As a check, the Schwarzschild solution is recovered through
these EFT equations from the Einstein--Hilbert Lagrangian, rewritten in
terms of the embedding and metric scalars. For comparison, in Appendix C we
present the alternative set of equations of motion obtained in the basis
adapted to the temporal foliation.

In Sec. VI we give the equations of motion for spherically symmetric, static
backgrounds in terms of generic Horndeski functions $G_{2}\left( \phi
,X\right) $ and $G_{4}\left( \phi \right) $ (but vanishing $G_{3}$ and $%
G_{5} $) and prove the unicity theorem that no action beyond
Einstein--Hilbert allows for the Schwarzschild solution in this class. With
this, we generalize the previously announced unicity theorem \cite{Sotiriou}
for the case of spherical symmetry and staticity.

In Sec. VII we discuss the EFT field equations for the case with only one
independent metric function and we formally integrate them to obtain the
metric function in terms of the nonvanishing Horndeski function $G_{4}\left(
\phi \right) $.

In Sec. VIII we obtain new solutions characterized by a parameter, which can
be interpreted as mass or tidal charge in the simplest particular cases, the
cosmological constant and a third parameter, emerging for various choices of 
$G_{4}\left( \phi \right) $. These solutions represent naked singularities,
hairy black holes or have the double horizon structure of the
Schwarzschild--de Sitter spacetime. Solutions with homogeneous
Kantowski--Sachs type regions also emerge.

In Sec. IX we rewrite these solutions as the conformally related metrics in
the Einstein frame.

Finally we give the concluding remarks in Sec. X.

\section{Nonorthogonal 2+1+1 spacetime decomposition: a summary}

This section summarizes the quantities and notations applied in the
nonorthogonal 2+1+1 decomposition of doubly foliable spacetimes worked out
in Ref. \cite{GKG}. The spacetime is foliated by 3-dimensional spacelike $%
\mathfrak{M}_{\chi }$ ($\chi =$const.) and timelike $\mathcal{S}_{t}$ ($t=$%
const.) hypersurfaces, whose 2-dimensional intersection is $\Sigma _{t\chi }$%
. The 4-dimensional metric can be decomposed as%
\begin{eqnarray}
\tilde{g}_{ab} &=&-n_{a}n_{b}+m_{a}m_{b}+g_{ab}  \notag \\
&=&-k_{a}k_{b}+l_{a}l_{b}+g_{ab}~.  \label{gabdecomp}
\end{eqnarray}%
Here $n_{a}$ and $l_{a}$ are normals to $\mathcal{S}_{t}$ and $\mathfrak{M}%
_{\chi }$, respectively. The metric tensor on $\Sigma _{t\chi }$ is $g_{ab}$%
, while the 1-form $m_{a}$ ($k_{a}$) is perpendicular to both $\Sigma
_{t\chi }$ and $n_{a}$ ($l_{a}$).

The tangent vectors of the coordinate\ lines $t$ and $\chi $ in the ($n^{a}$,%
$m^{a}$) basis can be given as 
\begin{eqnarray}
\left( \frac{\partial }{\partial t}\right) ^{a} &=&Nn^{a}+N^{a}+\mathcal{N}%
m^{a}~,  \label{ddt} \\
\left( \frac{\partial }{\partial \chi }\right) ^{a} &=&Mm^{a}+M^{a}~.
\label{ddchi}
\end{eqnarray}%
Here $N$ is the lapse function while $N^{a}$ (obeying $N^{a}m_{a}=0$) and $%
\mathcal{N}$ are the components of the 3-dimensional shift vector. In
addition $M$ is the lapse function of $\partial /\partial \chi $ in $%
\mathcal{S}_{t}$ and $M^{a}$ is its 2-dimensional shift vector which is
tangent to $\Sigma _{t\chi }$. The decompositions of $\partial /\partial t$
and $\partial /\partial \chi $ in the ($k^{a}$,$l^{a}$) basis are 
\begin{eqnarray}
\left( \frac{\partial }{\partial t}\right) ^{a} &=&\frac{N}{\mathfrak{c}}%
k^{a}+N^{a}~,  \label{ddtg} \\
\left( \frac{\partial }{\partial \chi }\right) ^{a} &=&M\left( -\mathfrak{s}%
k^{a}+\mathfrak{c}l^{a}\right) +M^{a}~,  \label{ddchig}
\end{eqnarray}%
where $\mathfrak{s}=\sinh \mathfrak{\psi }$, $\mathfrak{c}=\cosh \mathfrak{%
\psi }$ are defined by the Lorentz-rotation between the two bases, 
\begin{equation}
\begin{pmatrix}
k^{a} \\ 
l^{a}%
\end{pmatrix}%
=%
\begin{pmatrix}
\mathfrak{c} & \mathfrak{s} \\ 
\mathfrak{s} & \mathfrak{c}%
\end{pmatrix}%
\begin{pmatrix}
n^{a} \\ 
m^{a}%
\end{pmatrix}%
~,  \label{klnm}
\end{equation}%
with rapidity $\mathfrak{\psi }$ obeying 
\begin{equation}
\mathcal{N}=N\tanh \mathfrak{\psi }~.  \label{Phi}
\end{equation}

The covariant derivatives of the normals $n^{a}$, $l^{a}$ to the
hypersurfaces are decomposed in their naturally associated bases as%
\begin{eqnarray}
\tilde{\nabla}_{a}n_{b} &=&K_{ab}+2m_{(a}\mathcal{K}_{b)}+m_{a}m_{b}\mathcal{%
K}+n_{a}m_{b}\mathcal{L}^{\ast }  \notag \\
&&-n_{a}D_{b}\left( \ln N\right) ~,  \label{nfelb} \\
\tilde{\nabla}_{a}l_{b} &=&L_{ab}+2k_{(a}\mathcal{L}_{b)}+k_{a}k_{b}\mathcal{%
L}+l_{a}k_{b}\mathcal{K}^{\ast }  \notag \\
&&-l_{a}D_{b}\left( \ln \mathfrak{c}M\right) ~.  \label{lfelb}
\end{eqnarray}%
Here we have introduced $D$-derivatives representing\ covariant derivatives
of any 4-dimensional\ tensor $\tilde{T}_{b_{1}...b_{q}}^{a_{1}...a_{r}}$
projected onto $\Sigma _{t\chi }$ as 
\begin{equation}
D_{a}\tilde{T}_{b_{1}...b_{q}}^{a_{1}...a_{r}}\equiv
g_{a}^{c}g_{c_{1}}^{a_{1}}...g_{c_{r}}^{a_{r}}g_{b_{1}}^{d_{1}}...g_{b_{q}}^{d_{q}}%
\tilde{\nabla}_{c}\tilde{T}_{d_{1}...d_{q}}^{c_{1}...c_{r}}~.
\label{projkovd}
\end{equation}%
The 2-tensors $K_{ab}\equiv D_{a}n_{b}$ and $L_{ab}\equiv D_{a}l_{b}$ are
the extrinsic curvatures of $\Sigma _{t\chi }$, while 
\begin{eqnarray}
\mathcal{K}_{a} &\equiv &g_{a}^{c}m^{d}\tilde{\nabla}_{c}n_{d}=g_{a}^{c}m^{d}%
\tilde{\nabla}_{d}n_{c}~,  \notag \\
\mathcal{L}_{a} &\equiv &-g_{a}^{c}k^{d}\tilde{\nabla}%
_{c}l_{d}=-g_{a}^{c}k^{d}\tilde{\nabla}_{d}l_{c}=\mathcal{K}_{a}+D_{a}%
\mathfrak{\psi }~  \label{KLform}
\end{eqnarray}%
are normal fundamental forms, and 
\begin{eqnarray}
\mathcal{K} &\equiv &m^{d}m^{c}\tilde{\nabla}_{c}n_{d}~,  \notag \\
\mathcal{L} &\equiv &k^{d}k^{c}\tilde{\nabla}_{c}l_{d}~,  \label{KLscalar}
\end{eqnarray}%
the normal fundamental scalars defined by the hypersurface-orthogonal
vectors $n^{a}$ and $l^{a}$. The corresponding quantities for the basis
vectors $k^{a}$ and $m^{a}$ are\footnote{%
The embedding variables of the $\Sigma _{t\chi }$ surface with respect to
the normals of the hypersurfaces $n^{a}$, $l^{a}$, and those with respect to
the complementary orthogonal basis vectors $m^{a}$, $k^{a}$ will be
distinguished by a star on the latter set.}%
\begin{eqnarray}
\mathcal{K}^{\ast } &\equiv &l^{d}l^{c}\tilde{\nabla}_{c}k_{d}~,  \notag \\
\mathcal{L}^{\ast } &\equiv &n^{c}n^{d}\tilde{\nabla}_{c}m_{d}~.
\label{KLstarscalar}
\end{eqnarray}%
The vectorial and tensorial embedding variables generate additional scalars:%
\begin{eqnarray}
\varkappa &\equiv &K^{ab}K_{~ab}~,\quad \lambda \equiv L^{ab}L_{ab}~,  \notag
\\
\mathfrak{K} &\equiv &\mathcal{K}^{a}\mathcal{K}_{a}~,\quad \mathfrak{k}%
\equiv \mathcal{L}^{a}\mathcal{L}_{a}~,  \notag \\
K &\equiv &K_{~a}^{a}~,\quad L\equiv L_{~~a}^{a}~.  \label{scalars1}
\end{eqnarray}

The covariant derivatives of the complementary basis vectors $m^{a}$, $k^{a}$
can also be decomposed in their naturally associated bases,%
\begin{eqnarray}
\tilde{\nabla}_{a}k_{b} &=&K_{ab}^{\ast }+l_{a}\mathcal{K}_{b}^{\ast }+l_{b}%
\mathcal{L}_{a}+l_{a}l_{b}\mathcal{K}^{\ast }+k_{a}l_{b}\mathcal{L}  \notag
\\
&&-k_{a}D_{b}\left( \ln \frac{N}{\mathfrak{c}}\right) ~,  \label{kfelb} \\
\tilde{\nabla}_{a}m_{b} &=&L_{ab}^{\ast }+n_{a}\mathcal{L}_{b}^{\ast }+n_{b}%
\mathcal{K}_{a}+n_{a}n_{b}\mathcal{L}^{\ast }+m_{a}n_{b}\mathcal{K}  \notag
\\
&&+m_{a}D_{b}\left( \ln M\right) ~,  \label{mfelb}
\end{eqnarray}%
in terms of the extrinsic curvatures $K_{ab}^{\ast }\equiv D_{a}k_{b}$, $%
L_{ab}^{\ast }\equiv D_{a}m_{b}$ and quantities defined in a similar manner
to the normal fundamental forms, 
\begin{eqnarray}
\mathcal{K}_{a}^{\ast } &\equiv &g_{a}^{d}l^{c}\tilde{\nabla}_{c}k_{d}~, 
\notag \\
\mathcal{L}_{a}^{\ast } &\equiv &-g_{a}^{d}n^{c}\tilde{\nabla}_{c}m_{d}~.
\label{KLstarform}
\end{eqnarray}%
Scalars can be formed once again from the embedding tensors and vectors as%
\begin{eqnarray}
\varkappa ^{\ast } &\equiv &K^{\ast ab}K_{ab}^{\ast }~,\quad \lambda ^{\ast
}\equiv L^{\ast ab}L_{ab}^{\ast }~,  \notag \\
\mathfrak{K}^{\ast } &\equiv &\mathcal{K}^{\ast a}\mathcal{K}_{a}^{\ast
}~,\quad \mathfrak{k}^{\ast }\equiv \mathcal{L}^{\ast a}\mathcal{L}%
_{a}^{\ast }~,  \notag \\
K^{\ast } &\equiv &K_{~a}^{\ast a}~,\quad L^{\ast }\equiv L_{~~a}^{\ast a}~.
\label{scalars2}
\end{eqnarray}%
The scalars introduced for the two bases are interconnected; the starry ones
can be expressed in terms of their unstarred versions together with the
metric functions and their derivatives, as shown in Appendix \ref{App0}.

\section{Nonorthogonal 2+1+1 decomposition of the Horndeski action\label%
{Horndeski}}

We assume that the scalar has nowhere vanishing gradient and it solely
depends on $\chi $; hence, it defines a foliation through the $\phi $=const
level hypersurfaces. Hence the normal to the $\chi $=const hypersurfaces can
be expressed in terms of the scalar field gradient,

\begin{equation}
l_{a}=\frac{\tilde{\nabla}_{a}\phi }{\sqrt{X}}\,,  \label{lphigrad}
\end{equation}%
where $X=\tilde{g}^{ab}\tilde{\nabla}_{a}\phi \tilde{\nabla}_{b}\phi $ is
the kinetic term of the scalar field.

Time evolution along $\partial /\partial t$ proceeds on the $\chi $=const
hypersurfaces. In a perturbational setup this can be insured both on the
background and after perturbation, by absorbing the scalar field variation
into a proper gauge choice, dubbed radial unitary gauge \cite{KGT,GKG}.

Although for discussing the general relativistic Hamiltonian dynamics the
basis $\left( n^{a},m^{a}\right) $ turned more advantageous, as fewer
embedding variables were related to time derivatives of the metric
components \cite{GKG}, Eq. (\ref{lphigrad}) clearly indicates that for the
purpose of monitoring spherically symmetric, static configurations in
scalar-tensor gravity the $\left( k^{a},l^{a}\right) $ basis is better
suited.

Hence we proceed in rewriting the second covariant derivative of $\phi $ in
the $\left( k^{a},l^{a}\right) $ basis, 
\begin{gather}
\tilde{\nabla}_{a}\tilde{\nabla}_{b}\phi =\frac{\tilde{\nabla}^{c}\phi 
\tilde{\nabla}_{c}X}{2X}l_{a}l_{b}+\sqrt{X}\left[ L_{ab}+\mathcal{L}%
k_{a}k_{b}\right.  \notag \\
\left. +2\mathcal{K}^{\ast }l_{(a}k_{b)}+2k_{(a}\mathcal{L}%
_{b)}-2l_{(a}D_{b)}\ln \left( \mathfrak{c}M\right) \right] ~.
\label{nablanablaphi}
\end{gather}%
This generalizes Eq. (4.2) of Ref. \cite{KGT} for the case of nonorthogonal
double foliation.

The Horndeski action is%
\begin{equation}
S^{\mathrm{H}}=\int d^{4}x\sqrt{-\tilde{g}}L^{\mathrm{H}}~,
\end{equation}%
where the Lagrangian, 
\begin{equation}
L^{\mathrm{H}}=\sum_{i=2}^{5}L_{i}^{\mathrm{H}}\,,
\end{equation}%
is a sum of the contributions 
\begin{equation}
L_{2}^{\mathrm{H}}=G_{2}(\phi ,X)\,,  \label{L2}
\end{equation}%
providing scenarios with accelerated expansion, 
\begin{equation}
L_{3}^{\mathrm{H}}=G_{3}(\phi ,X)\tilde{\square}\phi \,,
\end{equation}%
enabling the Vainshtein screening-mechanism,%
\begin{eqnarray}
&&L_{4}^{\mathrm{H}}=G_{4}(\phi ,X)\tilde{R}-2G_{4X}(\phi ,X)  \notag \\
&&~~~~~~\times ~\left[ (\tilde{\square}\phi )^{2}\!-\!\tilde{\nabla}^{a}%
\tilde{\nabla}^{b}\phi \tilde{\nabla}_{a}\tilde{\nabla}_{b}\phi \right] \,,
\end{eqnarray}%
the first term of which leads to a time-evolving gravitational constant and
the much more involved $L_{5}^{\mathrm{H}}$, which has been disruled by both
the requirement of a working gravitational screening \cite%
{Vainshtein1,Vainshtein2,Vainshtein3} and the observation of the propagation
of gravitational waves equaling the speed of light \cite{Multimessenger}. In
fact $L_{4}^{\mathrm{H}}$ was also simplified by the latter observation,
disallowing the $X\,$-dependence of $G_{4}$. Nevertheless, we keep this
dependence for the time being as it does not add considerable difficulty to
our calculations.

Now we need to 2+1+1 decompose the respective parts of the Horndeski
Lagrangian and rewrite them in terms of the variables employed in the
decomposition. The trace of Eq. (\ref{nablanablaphi}) gives the d'Alembertian%
\begin{equation}
\tilde{\square}\phi =\frac{\tilde{\nabla}^{a}\phi \tilde{\nabla}_{a}X}{2X}+%
\sqrt{X}\left( L-\mathcal{L}\right) ~  \label{DAlembertphi}
\end{equation}%
Now $L_{3}^{\mathrm{H}}$ can be written in a more convenient form following
Ref. \cite{GLPV} by taking%
\begin{equation}
G_{3}(\phi ,X)=F_{3}(\phi ,X)+2XF_{3X}(\phi ,X)~,
\end{equation}%
then integrating $F_{3}\tilde{\square}\phi $ by parts and using (\ref%
{DAlembertphi}) in the term $2XF_{3X}\tilde{\square}\phi $. Then $L_{3}^{%
\mathrm{H}}$ reduces to the sum of 
\begin{equation}
L_{3}^{\mathrm{H}^{\prime }}=2X^{3/2}F_{3X}\left( L-\mathcal{L}\right)
-F_{3\phi }X~  \label{L3}
\end{equation}%
(the same as for orthogonal double foliation in Ref. \cite{KGT}) and a
covariant 4-divergence, to be dropped.

The curvature scalar $\tilde{R}$ appearing in $L_{4}^{\mathrm{H}}$ has to be
also 2+1+1 decomposed in terms of embedding variables defined in the $\left(
k^{a},l^{a}\right) $ basis. A similar lengthy derivation to the one employed
in deriving Eq. (52) of Ref. \cite{GKG} for the $\left( n^{a},m^{a}\right) $
basis, here yields 
\begin{eqnarray}
\tilde{R} &=&R+K^{\ast ab}K_{ab}^{\ast }-L^{ab}L_{ab}+2\mathcal{L}^{a}%
\mathcal{L}_{a}-K^{\ast }\left( K^{\ast }+2\mathcal{K}^{\ast }\right)  \notag
\\
&&+L\left( L-2\mathcal{L}\right) +2D^{a}\left( \ln \frac{N}{\mathfrak{c}}%
\right) D_{a}\ln \left( \mathfrak{c}M\right)  \notag \\
&&-2\tilde{\nabla}_{a}\left[ D^{a}\left( \ln NM\right) -\left( K^{\ast }+%
\mathcal{K}^{\ast }\right) k^{a}+\left( L-\mathcal{L}\right) l^{a}\right] ~,
\notag \\
&&  \label{Rkl}
\end{eqnarray}%
which still contains a covariant 4-divergence. One can get rid of it by
partially integrating in $L_{4}^{\mathrm{H}}$ (generating $G_{4\phi }$ and $%
G_{4X}$ terms), dropping the emerging covariant 4-divergences in the
process, obtaining 
\begin{eqnarray}
L_{4}^{\mathrm{H}^{\prime }} &=&G_{4}\left( R+K^{\ast ab}K_{ab}^{\ast
}-K^{\ast 2}\right) +2\sqrt{X}G_{4\phi }\left( L-\mathcal{L}\right)  \notag
\\
&&-\left( G_{4}-2XG_{4X}\right) \left[ L^{ab}L_{ab}-2\mathcal{L}^{a}\mathcal{%
L}_{a}+2K^{\ast }\mathcal{K}^{\ast }\right.  \notag \\
&&\left. -L^{2}+2L\mathcal{L}-2\left( D^{a}\ln \frac{N}{\mathfrak{c}}\right)
D_{a}\ln \left( \mathfrak{c}M\right) \right] ~.
\end{eqnarray}%
This can be rewritten in terms of the scalars defined in Eqs. (\ref{scalars1}%
), (\ref{scalars2}) as%
\begin{eqnarray}
L_{4}^{\mathrm{H}^{\prime }} &=&G_{4}\left( R+\varkappa ^{\ast }-K^{\ast
2}\right) +2\sqrt{X}G_{4\phi }\left( L-\mathcal{L}\right)  \notag \\
&&-\left( G_{4}-2XG_{4X}\right) \left[ \lambda -2\mathfrak{k}+2K^{\ast }%
\mathcal{K}^{\ast }-L^{2}\right.  \notag \\
&&\left. +2L\mathcal{L}-2\left( D^{a}\ln \frac{N}{\mathfrak{c}}\right)
D_{a}\ln \left( \mathfrak{c}M\right) \right] ~.  \label{Hornkl}
\end{eqnarray}

With this we have almost completed the program of rewriting the Horndeski
Lagrangian (without $L_{5}^{\mathrm{H}}$) in a form containing exclusively
scalars formed from the metric and embedding variables. The Lagrangian
depends on the metric variables $\left( \mathcal{N},N,M\right) $ and the
embedding scalars $\left( \mathcal{K}^{\ast },\mathfrak{k},K^{\ast
},\varkappa ^{\ast },\mathcal{L},L,\lambda \right) $ related to the $\left(
k,l\right) $ basis. It also depends on the scalar $\phi $, through the
unspecified functions $G_{2}$, $F_{3}$ , $G_{4}$ and through $X$. We address
the $X$ dependence of the Lagrangian by employing the expression of the
inverse metric (B3) of Ref. \cite{s+1+1a} in the coordinate basis. As in the
radial unitary gauge $\phi =\phi \left( \chi \right) $, only $\tilde{g}%
^{\chi \chi }=\left( N^{2}-\mathcal{N}^{2}\right) /N^{2}M^{2}=\left( 
\mathfrak{c}M\right) ^{-2}$ matters in calculating%
\begin{equation}
X=\tilde{g}^{\chi \chi }\left( \partial _{\chi }\phi \right) ^{2}=\left( 
\frac{\partial _{\chi }\phi }{\mathfrak{c}M}\right) ^{2}~.  \label{X}
\end{equation}

Beside the embedding variables the induced curvature scalar, calculated from
the induced metric $g_{ab}$, also appears in the formalism. We can assume
that $g_{ab}$ is diagonal, as all 2-metrics can be diagonalized. Further,
such a\ metric is locally conformally flat. Globally this might not be the
case as there may be singular points of the conformal factor. Hence for
spherical symmetry we take any 2-metric conformal to the unit sphere, while
for cylindrical symmetry to the plane, in both cases with conformal factor $%
\exp \left( 2\zeta \right) $. The two-metric hence is expressible by a
conformal exponent $\zeta $ alone. On the other hand, the 2-dimensional
curvature tensor built from this metric has only one independent component,
the curvature scalar $R$, also expressible in terms of the conformal
exponent. In summary, the Horndeski Lagrangian (without $L_{5}^{\mathrm{H}}$%
) only depends on the generalized coordinates 
\begin{equation}
\mathcal{N},N,M,\mathcal{K}^{\ast },\mathfrak{k},K^{\ast },\varkappa ^{\ast
},\mathcal{L},L,\lambda ,\zeta ,\phi ~  \label{variables}
\end{equation}%
and their derivatives.

On a technical note, the Horndeski Lagrangian derived above is much simpler,
than the one derivable in the $\left( n^{a},m^{a}\right) $ basis, which is
presented in the Appendix \ref{App1}, confirming the rightness of our basis
choice.

\section{The first order variation of the EFT action on spherically
symmetric, static background}

Having the Horndeski action rewritten in terms of the variables (\ref%
{variables}) our aim now is to derive the equations of motion by taking the
perturbations of the variables to first order about a background. We
summarize the procedure to be followed below. Quantities on the background
will be denoted by an overbar.

\subsection{First order variation of the action}

The variation of the action $S=\int d^{4}x\sqrt{-\tilde{g}}L$ to first order
formally reads 
\begin{equation}
\delta S=\int d^{4}x\overline{\sqrt{-\tilde{g}}}\overline{\frac{\delta S}{%
\delta G_{A}}}\delta G_{A}~,
\end{equation}%
where $\delta G_{A}\equiv G_{A}-\bar{G}_{A}$ for any variable $G_{A}$
(either related to the metric and scalar field) on which the action
functional depends and $\overline{\delta S/\delta G_{A}}~$is the functional
derivative of the action with respect to $G_{A}$, evaluated on the
background. For a first differential order dependence of the action
in\thinspace $G_{A}$ it can be expressed as the Euler--Lagrange expression, 
\begin{equation}
\overline{\sqrt{-\tilde{g}}}\overline{\frac{\delta S}{\delta G_{A}}}=%
\overline{\frac{\partial \left( \sqrt{-\tilde{g}}L\right) }{\partial G_{A}}}%
-\partial _{a}\overline{\frac{\partial \left( \sqrt{-\tilde{g}}L\right) }{%
\partial \left( \partial _{a}G_{A}\right) }}~.  \label{EL}
\end{equation}

As $\overline{\sqrt{-\tilde{g}}}$ does not depend on the derivatives of any
variable, 
\begin{eqnarray}
\frac{1}{\overline{\sqrt{-\tilde{g}}}}\partial _{a}\overline{\frac{\partial
\left( \sqrt{-\tilde{g}}L\right) }{\partial \left( \partial _{a}G_{A}\right) 
}} &=&\frac{1}{\overline{\sqrt{-\tilde{g}}}}\partial _{a}\left( \overline{%
\sqrt{-\tilde{g}}}\overline{\frac{\partial L}{\partial \left( \partial
_{a}G_{A}\right) }}\right)  \notag \\
&=&\overline{\tilde{\nabla}_{a}\frac{\partial L}{\partial \left( \partial
_{a}G_{A}\right) }}
\end{eqnarray}%
and the functional derivative (\ref{EL}) can be further expanded as 
\begin{equation}
\overline{\frac{\delta S}{\delta G_{A}}}=\overline{\frac{\partial L}{%
\partial G_{A}}}-\overline{\tilde{\nabla}_{a}\frac{\partial L}{\partial
\left( \partial _{a}G_{A}\right) }}+\bar{L}\overline{\frac{\partial \ln 
\sqrt{-\tilde{g}}}{\partial G_{A}}}~.
\end{equation}%
This enables us to write the generic form of the variation of the action,

\begin{equation}
\delta S=\int d^{4}x\sqrt{-\tilde{g}}\left( \,\delta L+L\delta \ln \sqrt{-%
\tilde{g}}\right) ~,  \label{varS}
\end{equation}%
with%
\begin{equation}
\delta L=\left( \overline{\frac{\partial L}{\partial G_{A}}}-\overline{%
\tilde{\nabla}_{a}\frac{\partial L}{\partial \left( \partial
_{a}G_{A}\right) }}\right) \delta G_{A}~,  \label{deltaL}
\end{equation}%
and 
\begin{equation}
\delta \ln \sqrt{-\tilde{g}}=\overline{\frac{\partial \left( \ln \sqrt{-%
\tilde{g}}\right) }{\partial G_{A}}}\delta G_{A}~.  \label{deltalng}
\end{equation}%
We will employ the expressions (\ref{varS})--(\ref{deltalng}) in what
follows.

On a spherically symmetric, static background we replace $\chi $ by the
radial coordinate $r$ and we denote the derivatives with respect to $r$ by a
prime.

When rewriting the Horndeski Lagrangian into the 2+1+1 formalism, most of
the variables appeared only algebraically. Even if the action depends on the
derivatives of the variables, hence the last term of Eq. (\ref{deltaL}) is
nonvanishing, when evaluated on the spherically symmetric and static
background, only the $r$-dependence survives; hence, 
\begin{equation}
\delta L=\left( \overline{\frac{\partial L}{\partial G_{A}}}-\overline{%
\tilde{\nabla}_{r}\frac{\partial L}{\partial \left( \partial
_{r}G_{A}\right) }}~\right) \delta G_{A}~.  \label{ELSSSB}
\end{equation}%
In what follows we further explore the particularities of the spherically
symmetric, static background.

\subsection{Metric and embedding variables on the spherically symmetric,
static background}

On the spherically symmetric, static background the metric can be chosen as%
\begin{equation}
ds^{2}=-\bar{N}^{2}dt^{2}+\bar{M}^{2}dr^{2}+r^{2}\left( d\theta ^{2}+\sin
^{2}\theta d\varphi ^{2}\right) ~  \label{ds2h}
\end{equation}%
implying that the following relations hold%
\begin{equation}
\bar{N}^{a}=\bar{M}^{a}=\mathcal{\bar{N}}=0~.  \label{BKG}
\end{equation}%
The latter assures that the two bases coincide on the background and 
\begin{eqnarray}
\bar{n}_{a} &=&\bar{k}_{a}=\left( -\bar{N},0,0,0\right) ~,  \notag \\
\bar{m}_{a} &=&\bar{l}_{a}=\left( 0,\bar{M},0,0\right) ~,
\end{eqnarray}%
hence%
\begin{equation}
\mathcal{\bar{K}}^{\ast }=\mathfrak{\bar{k}}=\bar{K}^{\ast }=\bar{\varkappa}%
^{\ast }=0~.  \label{Kk}
\end{equation}%
Further,%
\begin{eqnarray}
\mathcal{\bar{L}} &=&-\frac{\partial _{r}\bar{N}}{\bar{M}\bar{N}}~,  \notag
\\
\bar{L} &=&\frac{1}{2\bar{M}}\bar{g}^{ab}\partial _{r}\bar{g}_{ab}=\frac{2}{%
\bar{M}r}~,  \notag \\
\bar{\lambda} &=&\frac{2}{\bar{M}^{2}r^{2}}~,  \label{hatter}
\end{eqnarray}%
also 
\begin{equation}
\bar{L}_{ab}=\frac{1}{2}\bar{L}\bar{g}_{ab}~  \label{Labbar}
\end{equation}%
hold. From the 2-dimensional metric expressed in polar coordinates, $\bar{g}%
_{ab}=r^{2}$diag$\left( 1,\sin ^{2}\theta \right) $, the conformal exponent
(relating it to the unit sphere) $\bar{\zeta}=\ln r$ is found, also the
simple expression, 
\begin{equation}
\bar{R}=\frac{2}{r^{2}}~  \label{Rbar}
\end{equation}%
of the 2-dimensional curvature scalar emerges.

\subsection{The effective field theory type action}

The perturbed 2-dimensional metric and the background metric are related to
each other through an infinitesimal conformal factor $\delta \zeta $ as%
\begin{equation}
g_{ab}=e^{2\delta \zeta }\bar{g}_{ab}~;
\end{equation}%
hence, the variation of the 4-dimensional metric and of the 2-dimensional
Ricci scalar, under the assumption of spherical symmetric background emerges
as 
\begin{equation}
\delta \ln \sqrt{-\tilde{g}}=2\delta \zeta +\delta \ln N+\delta \ln M~,
\label{ad6}
\end{equation}%
and%
\begin{equation}
\delta R=R-\bar{R}=-2\bar{R}\delta \zeta -2\bar{g}^{ab}\bar{D}_{a}\bar{D}%
_{b}\delta \zeta ~.  \label{deltaRic}
\end{equation}

Based on (\ref{ad6}) and assuming for $L^{EFT}$ the same functional
dependence holds as for the Horndeski Lagrangians, we arrive to the
following effective field theory (EFT)\ type action%
\begin{equation}
S^{EFT}\left[ \mathcal{N},N,M,\mathcal{K}^{\ast },\mathfrak{k},K^{\ast
},\varkappa ^{\ast },\mathcal{L},L,\lambda ,\zeta ,\phi \right]
\label{actiondepend}
\end{equation}%
which can be viewed as a low-energy approximation of a yet unknown quantum
gravity.

Compared to the action (3.1) of Ref. \cite{KGT} the EFT action (\ref%
{actiondepend}) does not contain the dependence of the variable $\mathfrak{M}%
=M^{a}M_{a}$, as it did not appear in the Horndeski action; nevertheless, it
includes an explicit dependence of the radial shift $\mathcal{N}$,
nonexistent in the formalism of Ref. \cite{KGT}. Further differences are
hidden in the definitions of the variables $\left( \mathcal{K}^{\ast },%
\mathfrak{k},K^{\ast },\varkappa ^{\ast }\right) $ which reduce to the
corresponding ones $\left( \mathcal{K},\mathfrak{K},K,\varkappa \right) $ of
Ref. \cite{KGT} in the orthogonal double foliation limit.

\section{The field equations on spherically symmetric, static background 
\label{firstO}}

From the definitions of $\mathfrak{k}$ and $\varkappa ^{\ast }$ given in
Eqs. (\ref{scalars1}) and (\ref{scalars2}), respectively, together with
their vanishing on the background the variations $\delta \mathfrak{k}$ and $%
\delta \varkappa ^{\ast }$are second order, to be dropped from the first
order variation leading to the equations of motion. The $X$-dependence (\ref%
{X}) of the Horndeski Lagrangians (\ref{L2}), (\ref{L3}) and (\ref{Hornkl})
brings in $\phi ^{\prime }$, also additional $M$- and $\mathcal{N}$-terms.
Furthermore, the Horndeski Lagrangians do not depend on the derivatives of
any of the remaining geometric variables with the exception of the $-2\left(
D^{a}\ln \frac{N}{\mathfrak{c}}\right) D_{a}\ln \left( \mathfrak{c}M\right) $
term in $L_{4}^{H^{\prime }}$, which is again second order on the background
(on which $\bar{N}$ and $\bar{M}$ have only radial dependence, hence $D_{a}N$
and $D_{a}M$ are each first order).

As a result, for the purpose of first order variation we can assume the%
\textbf{\ }Lagrangian in the form [we do not explore yet Eq. (\ref{deltaRic}%
)], 
\begin{equation}
L^{EFT}\left( \mathcal{N},N,M,\mathcal{K}^{\ast },K^{\ast },\mathcal{L}%
,L,\lambda ,R,\phi ,\phi ^{\prime }\right) ~,  \label{LEFTfirstorder}
\end{equation}%
with 
\begin{eqnarray}
\delta L^{EFT} &=&L_{\mathcal{N}}^{EFT}\delta \mathcal{N}+L_{N}^{EFT}\delta
N+L_{M}^{EFT}\delta M\,  \notag \\
&&+L_{\mathcal{K}^{\ast }}^{EFT}\delta \mathcal{K}^{\ast }+L_{\mathfrak{k}%
}^{EFT}\delta \mathfrak{k}+\,L_{K^{\ast }}^{EFT}\delta K^{\ast }  \notag \\
&&+\,L_{\varkappa ^{\ast }}^{EFT}\delta \varkappa ^{\ast }+L_{\mathcal{L}%
}^{EFT}\delta \mathcal{L}+L_{L}^{EFT}\delta L  \notag \\
&&+L_{\lambda }^{EFT}\delta \lambda +L_{R}^{EFT}\delta R+L_{\phi
}^{EFT}\delta \phi ~,  \label{varL}
\end{eqnarray}%
where for all geometric variables $\mathcal{G}_{B}$ we denote%
\begin{equation}
L_{\mathcal{G}_{B}}^{EFT}=\overline{\frac{\partial \left( L^{EFT}\right) }{%
\partial \mathcal{G}_{B}}}~
\end{equation}%
and for the scalar%
\begin{equation}
L_{\phi }^{EFT}=\overline{\frac{\partial \left( L^{EFT}\right) }{\partial
\phi }}-\frac{1}{\overline{\sqrt{-\tilde{g}}}}\partial _{r}\left( \overline{%
\sqrt{-\tilde{g}}\frac{\partial \left( L^{EFT}\right) }{\partial \phi
^{\prime }}}\right) ~.  \label{ELP}
\end{equation}%
These are nothing but the expansion coefficients from Eq. (\ref{ELSSSB}),
taking into account that there is no dependence on the derivatives of the
geometric variables $\mathcal{G}_{B}$ of the Lagrangian (\ref{LEFTfirstorder}%
).

Now we proceed to compute the variation of the action (\ref{varS}) with the
contributions (\ref{varL})--(\ref{ELP}). From Eqs. (\ref{scalars1}), (\ref%
{hatter}) and (\ref{Labbar}), to first order,%
\begin{equation}
\delta \lambda =\frac{2}{\bar{M}r}\delta L~;  \label{dlambdastar}
\end{equation}%
thus,%
\begin{equation}
\int d^{4}x\sqrt{-\tilde{g}}\left( L_{L}^{EFT}\delta L+L_{\lambda
}^{EFT}\delta \lambda \right) =\int d^{4}x\sqrt{\bar{g}}\bar{M}\bar{N}%
\mathcal{F}\delta L~,  \label{ad1}
\end{equation}%
where we have replaced $\sqrt{-\tilde{g}}$ by its background value, as $%
\delta L$ is already first order and the quantity, 
\begin{equation}
\mathcal{F}=L_{L}^{EFT}+\frac{2}{\bar{M}r}L_{\lambda }^{EFT}~,  \label{Fdef}
\end{equation}%
defined similarly as in Ref. \cite{KGT}, is evaluated on the background.

Next from Eqs. (\ref{lfelb}) and (\ref{hatter})%
\begin{eqnarray}
L &=&\tilde{\nabla}_{a}l^{a}+\mathcal{\bar{L}}+\delta \mathcal{L}~,  \notag
\\
\delta L &=&\left[ \tilde{\nabla}_{a}l^{a}-\frac{\partial _{r}\bar{N}}{\bar{M%
}\bar{N}}-\frac{2}{\bar{M}r}\right] +\delta \mathcal{L}~.  \label{dLdL}
\end{eqnarray}%
From Eqs. (\ref{ddtg})--(\ref{ddchig}) for any scalar $G$ its directional
derivative can be expressed as 
\begin{eqnarray}
l^{a}\tilde{\nabla}_{a}G &=&\left[ \frac{\mathfrak{s}}{N}\partial _{t}+\frac{%
1}{\mathfrak{c}M}\partial _{r}\right.  \notag \\
&&\left. -\left( \frac{1}{\mathfrak{c}M}M^{a}+\frac{\mathfrak{s}}{N}%
N^{a}\right) D_{a}\right] G~;
\end{eqnarray}%
thus,

\begin{equation}
l^{a}\tilde{\nabla}_{a}\mathcal{F}=\frac{1}{\mathfrak{c}M}\partial _{r}%
\mathcal{F}~.  \label{lF}
\end{equation}%
Also the expansion of $\mathfrak{c}$ on the background gives $\mathfrak{c}%
=\cosh \psi =\allowbreak 1+\psi ^{2}/2+O\left( \psi ^{4}\right) $; thus, one
can safely replace it by\thinspace $1$ in a first order calculation.
Inserting Eq. (\ref{dLdL}) in Eq. (\ref{ad1}), employing integration by
parts and Eq. (\ref{lF}) we obtain%
\begin{eqnarray}
&&\int d^{4}x\sqrt{-\tilde{g}}\mathcal{F}\delta L=  \notag \\
&&\int d^{4}x\left[ \sqrt{\bar{g}}\bar{M}\bar{N}\tilde{\nabla}_{a}\left( 
\mathcal{F}l^{a}\right) -\partial _{r}\left( \sqrt{\bar{g}}\bar{N}\mathcal{F}%
\right) \right]  \notag \\
&&+\int d^{4}x\sqrt{\bar{g}}\bar{M}\bar{N}\left( \frac{\partial _{r}\mathcal{%
F}}{\bar{M}^{2}}\delta M+\mathcal{F}\delta \mathcal{L}\right) ~.
\label{Fint0}
\end{eqnarray}%
Further as $\bar{l}^{a}=\left( 0,\bar{M}^{-1},0,0\right) $ and exploring 
\begin{eqnarray*}
\overline{\sqrt{-\tilde{g}}} &=&\sqrt{-\tilde{g}}\left( 1-\delta \ln \sqrt{-%
\tilde{g}}\right) ~, \\
\tilde{\nabla}_{a}\bar{l}^{a} &=&\frac{1}{\bar{M}}\left( \frac{2}{r}+\frac{%
\partial _{r}\bar{N}}{\bar{N}}\right) ~,
\end{eqnarray*}%
the variational term (\ref{Fint0}) can be rewritten as a total covariant
divergence and a sum of variations $\delta M,~\delta \mathcal{L}$ and $%
\delta \ln \sqrt{-\tilde{g}}$:%
\begin{eqnarray}
&&\int d^{4}x\sqrt{-\tilde{g}}\mathcal{F}\delta L=  \notag \\
&&\int d^{4}x\sqrt{-\tilde{g}}\tilde{\nabla}_{a}\left( \mathcal{F}\bar{l}%
^{a}\delta \ln \sqrt{-\tilde{g}}\right)  \notag \\
&&+\int d^{4}x\sqrt{\bar{g}}\bar{M}\bar{N}\left( \frac{\partial _{r}\mathcal{%
F}}{\bar{M}^{2}}\delta M+\mathcal{F}\delta \mathcal{L}\right)  \notag \\
&&-\int d^{4}x\partial _{r}\left( \sqrt{\bar{g}}\bar{N}\mathcal{F}\right)
\delta \ln \sqrt{-\tilde{g}}~.
\end{eqnarray}

From the relations among embedding variables and coordinate derivatives,
given as Eqs. (35) and (36) of Ref. \cite{GKG}, the variation $\delta 
\mathcal{L}$ can be expressed as 
\begin{equation}
\delta \mathcal{L}=-\frac{\partial _{t}\delta \mathcal{N}}{\bar{N}^{2}}-%
\frac{\partial _{r}\delta N}{\bar{M}\bar{N}}+\frac{\partial _{r}\bar{N}}{%
\bar{M}\bar{N}}\left( \frac{\delta N}{\bar{N}}+\frac{\delta M}{\bar{M}}%
\right) ~,
\end{equation}%
The variation $\delta K^{\ast }$ and $\delta \mathcal{K}^{\ast }$ are also
related, since 
\begin{eqnarray}
\int d^{4}x\sqrt{-\tilde{g}}\,L_{K^{\ast }}^{EFT}\delta K^{\ast } &=&\,\int
d^{4}x\sqrt{-\tilde{g}}\,\tilde{\nabla}_{a}\left( L_{K^{\ast
}}^{EFT}k^{a}\right)  \notag \\
&&-\int d^{4}x\sqrt{-\tilde{g}}\,L_{K^{\ast }}^{EFT}\delta \mathcal{K}^{\ast
}~.  \notag \\
&&
\end{eqnarray}%
In addition from Eq. (35) of Ref. \cite{GKG} we find 
\begin{equation}
\delta \mathcal{K}^{\ast }=\frac{\partial _{t}\delta M}{\bar{M}\bar{N}}~.
\label{calKstar}
\end{equation}%
These generate $\partial _{t}\left[ \sqrt{\bar{g}}\left( L_{\mathcal{K}%
^{\ast }}^{EFT}-L_{K^{\ast }}^{EFT}\right) \right] $ as the prefactor of $%
\delta M$ in the first order variation of the action, which vanishes on a
static background.

Finally, the variation the action with respect to the geometrical variables
takes the form,%
\begin{eqnarray}
\delta _{\mathcal{G}}S^{EFT} &=&\delta _{\mathcal{G}}S_{\left( B\right)
}^{EFT}+\int d^{4}x\sqrt{-\tilde{g}}\,\left\{ L_{\mathcal{N}}^{EFT}\delta 
\mathcal{N}\,\,\right.  \notag \\
&&+\left[ \bar{L}^{EFT}+\bar{N}L_{N}^{EFT}\right.  \notag \\
&&\left. +\frac{1}{\bar{M}}\left( \frac{2}{r}+\frac{\partial _{r}\bar{N}}{%
\bar{N}}+\partial _{r}\right) L_{\mathcal{L}}^{EFT}\right] \frac{\delta N}{%
\bar{N}}  \notag \\
&&+\left( \bar{L}^{EFT}+\bar{M}L_{M}^{EFT}\right.  \notag \\
&&\left. +\frac{\partial _{r}\bar{N}}{\bar{M}\bar{N}}L_{\mathcal{L}}^{EFT}-%
\frac{2\mathcal{F}}{\bar{M}r}\right) \frac{\delta M}{\bar{M}}  \notag \\
&&+2\left[ L^{EFT}-\frac{2}{r^{2}}L_{R}^{EFT}\right.  \notag \\
&&\left. \left. -\frac{1}{\bar{M}}\left( \frac{2}{r}+\frac{\partial _{r}\bar{%
N}}{\bar{N}}+\partial _{r}\right) \mathcal{F}\right] \delta \zeta \right\} ~,
\label{deltaSEFT}
\end{eqnarray}%
where 
\begin{eqnarray}
\delta _{\mathcal{G}}S_{\left( B\right) }^{EFT} &=&\int d^{4}x\sqrt{-\tilde{g%
}}\tilde{\nabla}_{a}\left[ \mathcal{F}\bar{l}^{a}\delta \ln \sqrt{-\tilde{g}}%
\right.  \notag \\
&&+L_{K^{\ast }}^{EFT}k^{a}+\left( L_{\mathcal{K}^{\ast }}^{EFT}-L_{K^{\ast
}}^{EFT}\right) k^{a}\frac{\delta M}{\bar{M}}  \notag \\
&&\left. -\left( \mathcal{F}+L_{\mathcal{L}}^{EFT}\right) \left( k^{a}\frac{%
\delta \mathcal{N}}{\bar{N}}+l^{a}\frac{\delta N}{\bar{N}}\right) \right]
\end{eqnarray}%
is a boundary term. In the derivation we have used that $\sqrt{\bar{g}}%
=r^{2}\sin \theta $ and%
\begin{equation}
\int_{0}^{\pi }d\theta \int_{0}^{2\pi }d\varphi \sqrt{\bar{g}}\bar{D}%
_{a}\left( \bar{g}^{ab}\bar{D}_{b}\delta \zeta \right) =0~.
\end{equation}%
(The integral being a covariant divergence can be transformed to another
integral on the boundary of the sphere, which is zero.) The field equations
arising from (\ref{deltaSEFT}) are%
\begin{equation}
L_{\mathcal{N}}^{EFT}=0~,  \label{EOM1}
\end{equation}%
\begin{equation}
\bar{L}^{EFT}+\bar{N}L_{N}^{EFT}+\frac{1}{\bar{M}}\left( \frac{2}{r}+\frac{%
\partial _{r}\bar{N}}{\bar{N}}+\partial _{r}\right) L_{\mathcal{L}}^{EFT}=0~,
\label{EOM2}
\end{equation}%
\begin{equation}
\bar{L}^{EFT}+\bar{M}L_{M}^{EFT}+\frac{\partial _{r}\bar{N}}{\bar{M}\bar{N}}%
L_{\mathcal{L}}^{EFT}-\frac{2\mathcal{F}}{\bar{M}r}=0~,  \label{EOM3}
\end{equation}%
\begin{equation}
\bar{L}^{EFT}-\frac{2}{r^{2}}L_{R}^{EFT}-\frac{1}{\bar{M}}\left( \frac{2}{r}+%
\frac{\partial _{r}\bar{N}}{\bar{N}}+\partial _{r}\right) \mathcal{F}=0~.
\label{EOM4}
\end{equation}

Finally, the variation of the action with respect to the scalar field
results in the Euler-Lagrange equation:%
\begin{equation}
\left( r^{2}\bar{M}\bar{N}L_{\phi ^{\prime }}^{EFT}\right) ^{\prime }=r^{2}%
\bar{M}\bar{N}L_{\phi }^{EFT}~.  \label{EOM5}
\end{equation}

We note that for the Horndeski Lagrangians (\ref{L2}), (\ref{L3}) and (\ref%
{Hornkl}), whenever an orthogonal double foliation is chosen on the
background, $\mathcal{N}$ becomes first order and $\mathfrak{c}$, containing
only $\mathcal{N}^{2}$, of second order. As the only place $\mathcal{N}$
enters these Lagrangians is through $\mathfrak{c}$, Eq. (\ref{EOM1}) becomes
trivial. Then the dynamical equations are only (\ref{EOM2})-(\ref{EOM5}).
Despite starting with a modified set of variables and employing a
nonorthogonal double foliation, Eqs. (\ref{EOM2})-(\ref{EOM4}) are identical
with those derived in the orthogonal double foliation \cite{KGT}. For\ more
generic, beyond Horndeski $L^{EFT}$, however, Eq. (\ref{EOM1}) may carry
information.

For comparison, Appendix C also enlists the equations of motion obtained in
the $\left( n,m\right) $ basis. There, the analog of Eq. (\ref{EOM1}) is
nontrivial. The other equations are quite similar, but expressed in the
complementary set of variables.

\subsection{Derivation of the Schwarzschild solution from the EFT form of
the Einstein--Hilbert action}

As a check of the equations derived we first derive the Schwarzschild
solution of general relativity. As in this case there is no scalar field,
Eq. (\ref{EOM5}) is trivially satisfied and in light of the closing remark
of the previous section Eq. (\ref{EOM1}) is also trivial. For Eqs. (\ref%
{EOM2})-(\ref{EOM4}) one needs the Einstein--Hilbert Lagrangian density $%
L^{EH}$ up to first order perturbations on static and spherically symmetric
background. The twice-contracted Gauss equation (\ref{Rkl}) in the $\left(
k,l\right) $ basis contains bilinear expressions of $K^{\ast },\mathcal{K}%
^{\ast }$, which due to Eq. (\ref{Kk}) are second order. We can also drop a
covariant four-divergence term; hence,%
\begin{equation}
L^{EH}=R-\lambda +L\left( L-2\mathcal{L}\right) ~,
\end{equation}%
which on the background gives%
\begin{equation}
\bar{L}^{EH}=\frac{2}{r^{2}}+\frac{2}{\bar{M}^{2}r^{2}}+\frac{4\bar{N}%
^{\prime }}{\bar{M}^{2}\bar{N}r}~.  \label{LEH}
\end{equation}%
Using Eqs. (\ref{hatter})--(\ref{Rbar}) and (\ref{Fdef}), the field
equations reduce to%
\begin{equation}
\bar{M}^{2}-1+2r\frac{\bar{M}^{\prime }}{\bar{M}}=0~,  \label{Schw1}
\end{equation}%
\begin{equation}
\bar{M}^{2}-1-2r\frac{\bar{N}^{\prime }}{\bar{N}}=0~,
\end{equation}%
\begin{equation}
\frac{r}{\bar{N}}\left( r\bar{N}^{\prime }\right) ^{\prime }-r^{2}\frac{\bar{%
M}^{\prime }}{\bar{M}}\left( \frac{1}{r}+\frac{\bar{N}^{\prime }}{\bar{N}}%
\right) =0~.  \label{Schw3}
\end{equation}%
The first two equations immediately give%
\begin{equation}
\bar{N}\propto \bar{M}^{-1}~.  \label{NMmone}
\end{equation}%
The proportionality coefficient can be chosen as $1$ by redefining the
coordinate $r$. Then the left-hand side of Eq. (\ref{Schw3}) can be
rewritten as%
\begin{equation}
\frac{r}{\bar{N}}\left( r\bar{N}^{\prime }\right) ^{\prime }+r^{2}\frac{\bar{%
N}^{\prime }}{\bar{N}}\left( \frac{1}{r}+\frac{\bar{N}^{\prime }}{\bar{N}}%
\right) =\frac{\left[ r^{2}\left( \bar{N}^{2}\right) ^{\prime }\right]
^{\prime }}{2\bar{N}^{2}}  \label{rewritelna}
\end{equation}%
such that Eq. (\ref{Schw3}) immediately gives 
\begin{equation}
\bar{N}^{2}=K\left( 1-\frac{C}{r}\right) ~.
\end{equation}%
The factor $K$ can be chosen as $1$ by redefining the time coordinate, and
the weak field limit leads to $C=2m$.

\section{EFT\ equations of motion for nonminimally coupled k-essence}

In this section we discuss the Horndeski theories with $G_{3}=G_{5}=0$,
generic $G_{2}\left( \phi ,X\right) $ (k-essence) and $G_{4}(\phi )$
(nonminimal coupling to the metric).

The Lagrangian density at first order in perturbations on static and
spherically symmetric background reduces to%
\begin{eqnarray}
L^{EFT} &=&G_{2}\left( \phi ,X\right) +G_{4}\left( \phi \right) \left(
R-\lambda +L^{2}-2L\mathcal{L}\right)  \notag \\
&&+2\sqrt{X}G_{4\phi }\left( \phi \right) \left( L-\mathcal{L}\right) ~.
\label{LfR}
\end{eqnarray}%
On the background 
\begin{eqnarray}
\bar{L}^{EFT} &=&\bar{G}_{2}+2\bar{G}_{4}\left( \frac{1}{r^{2}}+\frac{1}{%
\bar{M}^{2}r^{2}}+\frac{2\bar{N}^{\prime }}{\bar{M}^{2}\bar{N}r}\right) 
\notag \\
&&+\frac{2\phi ^{\prime }}{\bar{M}^{2}}\bar{G}_{4\phi }\left( \frac{2}{r}+%
\frac{\bar{N}^{\prime }}{\bar{N}}\right) ~.
\end{eqnarray}%
The nontrivial field equations (\ref{EOM2})--(\ref{EOM4}) are%
\begin{gather}
\bar{M}^{2}-1+2r\frac{\bar{M}^{\prime }}{\bar{M}}=\frac{r^{2}}{\bar{G}_{4}}%
\left[ -\frac{\bar{M}^{2}}{2}\bar{G}_{2}\right.  \notag \\
\left. +\left( \frac{2}{r}-\frac{\bar{M}^{\prime }}{\bar{M}}+\partial
_{r}\right) \left( \bar{G}_{4\phi }\phi ^{\prime }\right) \right] ~,
\label{Hspec1}
\end{gather}%
\begin{gather}
\bar{M}^{2}-1-2r\frac{\bar{N}^{\prime }}{\bar{N}}=\frac{r^{2}}{\bar{G}_{4}}%
\left[ -\frac{\bar{M}^{2}}{2}\bar{G}_{2}\right.  \notag \\
\left. +\left( \frac{2}{r}+\frac{\bar{N}^{\prime }}{\bar{N}}\right) \bar{G}%
_{4\phi }\phi ^{\prime }+\phi ^{\prime 2}\bar{G}_{2X}\right] ~,
\label{Hspec2}
\end{gather}%
and%
\begin{gather}
\frac{r}{\bar{N}}\left( r\bar{N}^{\prime }\right) ^{\prime }-r^{2}\frac{\bar{%
M}^{\prime }}{\bar{M}}\left( \frac{1}{r}+\frac{\bar{N}^{\prime }}{\bar{N}}%
\right) =\frac{r^{2}}{\bar{G}_{4}}\left[ \frac{\bar{M}^{2}}{2}\bar{G}%
_{2}\right.  \notag \\
\left. -\left( \frac{1}{r}+\frac{\bar{N}^{\prime }}{\bar{N}}-\frac{\bar{M}%
^{\prime }}{\bar{M}}+\partial _{r}\right) \left( \bar{G}_{4\phi }\phi
^{\prime }\right) \right] ~,  \label{Hspec3}
\end{gather}%
while the scalar equation (\ref{EOM5}) becomes%
\begin{gather}
\left( \frac{r^{2}\bar{N}}{\bar{M}}\phi ^{\prime }\bar{G}_{2X}\right)
^{\prime }-\frac{r^{2}\bar{M}\bar{N}}{2}\bar{G}_{2\phi }=\frac{\bar{N}}{\bar{%
M}}\left[ \bar{M}^{2}-1-\frac{2r\bar{N}^{\prime }}{\bar{N}}\right.  \notag \\
\left. +r^{2}\frac{\bar{M}^{\prime }}{\bar{M}}\left( \frac{2}{r}+\frac{\bar{N%
}^{\prime }}{\bar{N}}\right) -r^{2}\frac{\bar{N}^{\prime \prime }}{\bar{N}}%
\right] \bar{G}_{4\phi }~.  \label{Hspec4}
\end{gather}

For $G_{4}=\phi $ and $G_{2}=3X/2\phi -V\left( \phi \right) $ the system (%
\ref{LfR}) reproduces $f\left( R\right) $-gravity \cite{CapoFara}, while the
even more generic setup with $G_{4}=\phi $ and $G_{2}=-\omega \left( \phi
\right) X-V\left( \phi \right) $ was considered by Sotiriou and Faraoni \cite%
{Sotiriou}. Imposing asymptotic flatness and the vanishing of $V\left( \phi
\right) $ at infinity, also forbidding linear instabilities of the scalar in
the Einstein frame, they proved that i) the scalar field ought to be
constant (then the conformal factor is also constant and asymptotic flatness
also holds in the Einstein frame), ii) $V\left( \phi \right) =0$ holds in
the entire spacetime; therefore $G_{2}=0$. With these, the surviving
theories are described by the Einstein--Hilbert action.

The consistency of the EFT formalism discussed above can be also verified by
imposing the outcome of the Sotiriou--Faraoni unicity theorem, $G_{2}=0=\phi
^{\prime }$ leading to the Einstein-Hilbert action. With these conditions
the right-hand sides of Eqs. (\ref{Hspec1})-(\ref{Hspec4}) vanish, hence
they are the same as Eqs. (\ref{Schw1})-(\ref{Schw3}), leading immediately
to the Schwarzschild solution. Since $\phi ^{\prime }=0$ and $G_{2}=0$ the
left-hand side of the scalar equation (\ref{Hspec4}) vanishes. On the
right-hand side the coefficient of $\bar{G}_{4\phi }$ is a linear
combination of the left-hand sides of Eqs. (\ref{Hspec2}), (\ref{Hspec3})
and $\left( \bar{N}^{\prime }/\bar{N}+\bar{M}^{\prime }/\bar{M}\right) $.
For the Schwarzschild solution the first two obviously vanish, while the
third vanishes due to Eq. (\ref{NMmone}).

\subsection{Unicity theorem for the Schwarzschild solution}

The equations for the geometric variables (\ref{Hspec1})--(\ref{Hspec3})
reduce to the corresponding ones for the Schwarzschild solution, Eqs. (\ref%
{Schw1})-(\ref{Schw3}), whenever the conditions, 
\begin{equation}
\frac{\bar{M}^{2}}{2}\bar{G}_{2}=\left( \frac{2}{r}-\frac{\bar{M}^{\prime }}{%
\bar{M}}+\partial _{r}\right) \left( \bar{G}_{4\phi }\phi ^{\prime }\right)
~,  \label{HS1}
\end{equation}%
\begin{equation}
\frac{\bar{M}^{2}}{2}\bar{G}_{2}=\left( \frac{2}{r}+\frac{\bar{N}^{\prime }}{%
\bar{N}}\right) \bar{G}_{4\phi }\phi ^{\prime }+\phi ^{\prime 2}\bar{G}%
_{2X}~,  \label{HS2}
\end{equation}%
and%
\begin{equation}
\frac{\bar{M}^{2}}{2}\bar{G}_{2}=\left( \frac{1}{r}+\frac{\bar{N}^{\prime }}{%
\bar{N}}-\frac{\bar{M}^{\prime }}{\bar{M}}+\partial _{r}\right) \left( \bar{G%
}_{4\phi }\phi ^{\prime }\right) ~  \label{HS3}
\end{equation}%
hold. Taking the difference of Eqs. (\ref{HS3}) and (\ref{HS1}) one gets%
\begin{equation}
\left( \frac{\bar{N}^{\prime }}{\bar{N}}-\frac{1}{r}\right) \bar{G}_{4\phi
}\phi ^{\prime }=0~.
\end{equation}%
As for the Schwarzschild solution the first factor does not vanish, $\bar{G}%
_{4\phi }\phi ^{\prime }=\bar{G}_{4}^{\prime }=0$; hence $\bar{G}_{4}$ is a
constant. Then from Eq. (\ref{HS3}) $\bar{G}_{2}=0$ also follows.

With this, we proved the following unicity theorem. \textit{In the Horndeski
class of theories with generic }$G_{2}\left( \phi ,X\right) $\textit{\ and }$%
G_{4}\left( \phi \right) $\textit{\ functions (but }$G_{3}$\textit{\ and }$%
G_{5}$\textit{\ vanishing) only the Einstein-Hilbert action can allow for
the Schwarzschild solution.}

\section{A class of exact solutions for nonminimally coupled k-essence}

We continue our discussion in the framework of Horndeski theories with $%
G_{3}=G_{5}=0$, generic $G_{2}\left( \phi ,X\right) $ (k-essence) and $%
G_{4}(\phi )$ (nonminimal coupling to the metric) by imposing%
\begin{equation}
\bar{N}=\bar{M}^{-1}~,  \label{onefunc}
\end{equation}%
thus, allowing for only one undetermined metric function. With this choice
in the weak and stationary field limit there is only one potential appearing
in the metric. If that deviates only slightly from the Newtonian potential,
both Solar System and gravitational lensing tests of general relativity
could be reproduced. One can rewrite the line element in the
Eddington--Finkelstein forms%
\begin{eqnarray}
ds^{2}\! &=&\!-\!\bar{N}^{2}du^{2}\!-\!2dudr\!+\!r^{2}\left( d\theta
^{2}\!+\!\sin ^{2}\theta d\varphi ^{2}\right)  \notag \\
\! &=&\!-\!\bar{N}^{2}dv^{2}\!+\!2drdv\!+\!r^{2}\left( d\theta ^{2}\!+\!\sin
^{2}\theta d\varphi ^{2}\right) ~,
\end{eqnarray}%
with $r^{\ast }=\int dr/\bar{N}^{2}\left( r\right) $ the tortoise
coordinate, $u=t-r^{\ast }$ and $v=t+r^{\ast }$ as the retarded (outgoing)
and advanced (ingoing) time, respectively. For outgoing radial light rays in
the ingoing Eddington--Finkelstein coordinates $dr/dv=\bar{N}^{2}/2$; hence
at $\bar{N}=0$ there is an apparent and event horizon.

With the choice (\ref{onefunc}), Eqs. (\ref{Hspec1})--(\ref{Hspec4}) reduce
to%
\begin{gather}
1-\bar{N}^{2}-r\left( \bar{N}^{2}\right) ^{\prime }=\frac{r^{2}}{\bar{G}_{4}}%
\left[ -\frac{1}{2}\bar{G}_{2}\right.  \notag \\
\left. +\left( \frac{2\bar{N}^{2}}{r}+\frac{\left( \bar{N}^{2}\right)
^{\prime }}{2}+\bar{N}^{2}\partial _{r}\right) \left( \bar{G}_{4\phi }\phi
^{\prime }\right) \right] ~,  \label{Hspec1a}
\end{gather}%
\begin{gather}
1-\bar{N}^{2}-r\left( \bar{N}^{2}\right) ^{\prime }=\frac{r^{2}}{\bar{G}_{4}}%
\left[ -\frac{1}{2}\bar{G}_{2}\right.  \notag \\
\left. +\left( \frac{2\bar{N}^{2}}{r}+\frac{\left( \bar{N}^{2}\right)
^{\prime }}{2}\right) \bar{G}_{4\phi }\phi ^{\prime }+\bar{N}^{2}\phi
^{\prime 2}\bar{G}_{2X}\right] ~,  \label{Hspec2a}
\end{gather}%
\begin{gather}
\frac{\left[ r^{2}\left( \bar{N}^{2}\right) ^{\prime }\right] ^{\prime }}{2}=%
\frac{r^{2}}{\bar{G}_{4}}\left[ \frac{1}{2}\bar{G}_{2}\right.  \notag \\
\left. -\left( \frac{1}{r}\bar{N}^{2}+\left( \bar{N}^{2}\right) ^{\prime }+%
\bar{N}^{2}\partial _{r}\right) \left( \bar{G}_{4\phi }\phi ^{\prime
}\right) \right] ~,  \label{Hspec3a}
\end{gather}%
and%
\begin{gather}
\left( r^{2}\bar{N}^{2}\phi ^{\prime }G_{2X}\right) ^{\prime }-\frac{r^{2}}{2%
}\bar{G}_{2\phi }  \notag \\
=\left( 1-\bar{N}^{2}-r\left( \bar{N}^{2}\right) ^{\prime }-\frac{\left[
r^{2}\left( \bar{N}^{2}\right) ^{\prime }\right] ^{\prime }}{2}\right) \bar{G%
}_{4\phi }~.  \label{Hspec4a}
\end{gather}%
Both for the left-hand side of Eq. (\ref{Hspec3a}) and the right-hand side
of Eq. (\ref{Hspec4a}) we have explored Eq. (\ref{rewritelna}). By taking
the difference of Eqs. (\ref{Hspec1a}) and (\ref{Hspec2a}) the following
simple equation emerges:%
\begin{equation}
\left( \bar{G}_{4\phi }\phi ^{\prime }\right) ^{\prime }=\phi ^{\prime 2}%
\bar{G}_{2X}~.  \label{Hspec2b}
\end{equation}%
Taking the sum of Eqs. (\ref{Hspec1a}) and (\ref{Hspec3a}) and multiplying
by $\bar{G}_{4}$ one obtains an equation with only $\bar{G}_{4}$ and $\bar{N}
$,%
\begin{equation}
\left[ 1-\bar{N}^{2}-r\left( \bar{N}^{2}\right) ^{\prime }\right] \bar{G}%
_{4}+\left( \frac{r^{2}\left( \bar{N}^{2}\right) ^{\prime }}{2}\bar{G}%
_{4}\right) ^{\prime }=r\bar{N}^{2}\bar{G}_{4}^{\prime }~,  \label{Hspec3b}
\end{equation}%
which can be rewritten as%
\begin{equation}
\bar{G}_{4}=-\left[ \left( \frac{\bar{N}^{2}}{r^{2}}\right) ^{\prime }\frac{%
r^{4}\bar{G}_{4}}{2}\right] ^{\prime }~,
\end{equation}%
with the solution%
\begin{equation}
\bar{N}^{2}=-2r^{2}\int^{r}\frac{d\sigma }{\sigma ^{4}\bar{G}_{4}\left( \phi
\left( \sigma \right) \right) }\int^{\sigma }d\rho \bar{G}_{4}\left( \phi
\left( \rho \right) \right) ~.  \label{metric}
\end{equation}%
With this, the metric is fully determined in terms of $\bar{G}_{4}$ and two
integration constants. Then Eqs. (\ref{Hspec2b}) and (\ref{Hspec4a}) give $%
\bar{G}_{2X}$ and $\bar{G}_{2\phi }$, respectively, while $\bar{G}_{2}$
itself is given by Eq. (\ref{Hspec2a}).

\section{Black hole, naked singularity and homogeneous solutions\label{new}}

Let us explore a few particular cases of the general solution derived in the
previous section.

\subsection{Constant $\bar{G}_{4}$}

Assuming $\bar{G}_{4}=\left( 16\pi G\right) ^{-1}$ a constant, a minimal
coupling of constant scalar and metric is realized, hence the Einstein and
Jordan frames coincide. In this case the metric function becomes%
\begin{equation}
\bar{N}^{2}=1-\frac{2m}{r}-\Lambda r^{2}~.  \label{SchdS}
\end{equation}%
Here $m$ and $\Lambda $ are integration constants and we obtained the
Schwarzschild--de Sitter metric for $\Lambda >0$ and Schwarzschild--anti de
Sitter metric for $\Lambda <0$. Then Eq. (\ref{Hspec2b}) gives $\phi
^{\prime }\bar{G}_{2X}=0$. With $\phi ^{\prime }=0$ also $X=0\,$\ holds,
hence $\bar{G}_{2X}=0$. Next Eq. (\ref{Hspec4a}) yields $\bar{G}_{2\phi }=0$%
, yielding $\bar{G}_{2}$ to a constant, to be found from Eq. (\ref{Hspec2a})
as $\bar{G}_{2}=-6\Lambda /\left( 16\pi G\right) $. Thus $\bar{G}_{2}$
contributes a cosmological constant term to the action. This is why the
Schwarzschild--(anti) de Sitter metric emerged.\footnote{%
In another context, in the particular case of a constant $\bar{G}_{4}$ and%
\textbf{\ }$\bar{G}_{2}\left( X\right) $ Ref. \cite{GBDerivC7} also arrived
to a constant $\bar{G}_{2}$, leading to the Schwarzschild--(anti) de Sitter
metric.}

\subsection{The case $\bar{G}_{4}=\protect\phi =r$}

When both $\bar{G}_{4}$ and its inverse are regular, $\bar{G}_{4}$ can be
identified with the scalar \cite{Sotiriou}. If further, the scalar is a
monotonic function of the radial coordinate with a nowhere vanishing
derivative, it can be chosen as the radial coordinate itself. Here we
explore the case when the scalar is the curvature coordinate. The metric
function in this case becomes%
\begin{equation}
\bar{N}^{2}=\frac{1}{2}+\frac{Q}{r^{2}}-\Lambda r^{2}~.
\end{equation}%
Here $Q$ and $\Lambda $ are integration constants, interpreted as tidal
charge and cosmological constant. The metric has a curvature singularity in
the origin and evades asymptotic flatness due to the term $1/2$ even in the
absence of the cosmological constant, as%
\begin{equation}
\lim_{_{\Lambda =0}^{r\rightarrow \infty }}R_{\theta \varphi \theta }^{~\ \
\ \ \varphi }=\frac{1}{2}~,~\lim_{_{\Lambda =0}^{r\rightarrow \infty
}}R_{\theta \varphi \varphi }^{~\ \ \ \ \theta }=-\frac{\sin ^{2}\theta }{2}%
~.  \label{RiemannAtInfinity}
\end{equation}%
Equation (\ref{Hspec2b}) gives $\bar{G}_{2X}=0$. Next Eq. (\ref{Hspec4a})
yields $\bar{G}_{2\phi }=-12\Lambda -1/r^{2}$, which easily integrates to%
\begin{equation}
\bar{G}_{2}\left( \phi \right) =-12\Lambda \phi +\frac{1}{\phi }~.
\label{G21}
\end{equation}%
Equation (\ref{Hspec2a}) fixed the integration constant to zero.

\subsubsection{Horizons}

The metric function $\bar{N}^{2}$ vanishes for

\begin{equation*}
r_{1,2}^{2}=\frac{1\pm \sqrt{1+16Q\Lambda }}{4\Lambda }~,
\end{equation*}%
provided $16Q\Lambda \geq -1$. In this range for $\Lambda >0$ and $Q<0$
there are two positive roots, hence horizons. The metric function $\bar{N}%
^{2}$ is positive only between them. In this case the metric represents a
Kantowski-Sachs type homogeneous solution inside the smaller horizon, a
spherically symmetric, static black hole in between, while outside the
larger, cosmological horizon it is homogeneous again, asymptotically
approaching anti de Sitter.

For $\Lambda <0$ and $Q>0$ there is no horizon hiding the central
singularity.

For $\Lambda <0$ and $Q<0$ there is a positive root (hence horizon) at 
\begin{equation}
r_{1}=\frac{1}{2}\sqrt{\frac{1-\sqrt{1+16Q\Lambda }}{\Lambda }}.
\end{equation}%
Above this horizon $\bar{N}^{2}>0$ and below $\bar{N}^{2}<0$; hence the
solution represents a spherically symmetric, static black hole with
homogeneous interior.

Finally, for $\Lambda >0$ and $Q>0$ a horizon appears at%
\begin{equation}
r_{2}=\frac{1}{2}\sqrt{\frac{1+\sqrt{1+16Q\Lambda }}{\Lambda }}.
\end{equation}%
Then $\bar{N}^{2}>0$ holds below the anti de Sitter type cosmological
horizon.

\subsection{The case $\bar{G}_{4}=\protect\phi =r^{\protect\alpha }$}

The metric function becomes 
\begin{equation}
\bar{N}^{2}=\frac{1}{1+\alpha }+\frac{C}{r^{1+\alpha }}-\Lambda r^{2}~,
\label{Nalpha}
\end{equation}%
with $C$ and $\Lambda $ integration constants. This metric includes the
previous two cases for $\alpha =0,1$ (with $C=2m,$ $Q$ in these cases,
respectively). Here we allow all $\alpha \geq 0$; hence the metric is
singular in the origin. The metric has a curvature singularity in the origin
and evades asymptotic flatness when $\alpha \neq 0$, even in the absence of
the cosmological constant, as%
\begin{equation}
\lim_{_{\Lambda =0}^{r\rightarrow \infty }}R_{\theta \varphi \theta }^{~\ \
\ \ \varphi }=\frac{\alpha }{1+\alpha }~,~\lim_{_{\Lambda =0}^{r\rightarrow
\infty }}R_{\theta \varphi \varphi }^{~\ \ \ \ \theta }=-\frac{\alpha \sin
^{2}\theta }{1+\alpha }~.
\end{equation}%
Then Eq. (\ref{Hspec2b}) gives 
\begin{equation}
\bar{G}_{2X}=\frac{\alpha -1}{\alpha r^{\alpha }}~.
\end{equation}%
Next Eq. (\ref{Hspec2a}) gives%
\begin{equation}
\bar{G}_{2}=\frac{2\alpha ^{2}r^{\alpha -2}}{1+\alpha }+\frac{\alpha \left(
\alpha -1\right) C}{r^{3}}-2\left( 3+2\alpha +\alpha ^{2}\right) \Lambda
r^{\alpha }~.  \label{G2alpha}
\end{equation}%
In the particular case $\alpha =1$ Eq. (\ref{G21}) is recovered, while for $%
\alpha =0$ the desired cosmological constant type contribution to the action
emerges.

From Eq. (\ref{Hspec4a}) we find%
\begin{equation}
\bar{G}_{2\phi }=-\frac{2}{\left( 1+\alpha \right) r^{2}}-\frac{\alpha
\left( \alpha -1\right) C}{r^{\alpha +3}}-6\left( \alpha +1\right) \Lambda ~.
\label{G2phialpha}
\end{equation}%
With%
\begin{equation}
X=N^{2}\phi ^{\prime 2}=\alpha ^{2}\left( \frac{r^{2\alpha -2}}{1+\alpha }%
+Cr^{\alpha -3}-\Lambda r^{2\alpha }\right) ~,
\end{equation}%
Eq. (\ref{G2alpha}) can be rewritten in terms of the scalar field and
kinetic term as%
\begin{equation}
\bar{G}_{2}=\frac{\alpha -1}{\alpha }\frac{X}{\phi }+\alpha \phi ^{\frac{%
\alpha -2}{\alpha }}-\left( 6+5\alpha +\alpha ^{2}\right) \Lambda \phi ~,
\label{G2alphaphiX}
\end{equation}%
which correctly reproduces both $\bar{G}_{2X}$ and $\bar{G}_{2\phi }$. For $%
\alpha =1$ we recover Eq. (\ref{G21}) while for $\alpha =0$ (taking into
account that $X\propto \alpha ^{2}$ and for constant $G_{4}$ the scalar is
also a constant) the cosmological constant type contribution to the action
reemerges.

\subsubsection{Discussion of the horizons}

The location of the horizons is determined by%
\begin{equation}
-\Lambda r^{3+\alpha }+\frac{1}{\left( 1+\alpha \right) }r^{1+\alpha }+C=0~.
\end{equation}%
For $\Lambda =0$ and for $C<0$ there is one real solution, hence horizon at 
\begin{equation}
r=\sqrt[1+\alpha ]{-\left( 1+\alpha \right) C}~.
\end{equation}%
Similarly, for $C=0$ and $\Lambda >0$ a horizon can be found at 
\begin{equation}
r=\sqrt{\frac{1}{\left( 1+\alpha \right) \Lambda }}~.
\end{equation}%
For $\Lambda \neq 0$ and integer $\alpha $ the number of positive real roots
is given by the Descartes' rule of sign. According to it for $\Lambda $
negative and $C$ positive there is no horizon. When $\Lambda $ and $C$ have
identical signs, a horizon exists. When $\Lambda $ is positive and $C$
negative there is either no horizon or there are two horizons, according to
this rule.

\subsection{The linear case $\bar{G}_{4}=\protect\phi =A\left( 1+Br\right) $}

With two integrations the metric function (\ref{metric}) becomes%
\begin{eqnarray}
\bar{N}^{2} &=&1+3Bm-\frac{2m}{r}-B\left( 1+6Bm\right) r-\Lambda r^{2} 
\notag \\
&&-B^{2}\left( 1+6Bm\right) r^{2}\ln \left\vert \frac{Br}{1+Br}\right\vert ~,
\label{N2new}
\end{eqnarray}%
where $m$ and$~\Lambda $ are integration constants. The case $B=0$
reproduces the earlier found Schwarzshild--(anti) de Sitter metric.
Asymptotically the last term of Eq. (\ref{N2new}) vanishes and the solution
approaches (anti) de Sitter with the cosmological constant $\Lambda $. The
independent, nonvanishing components of the Riemann tensor at spatially
infinity for $\Lambda =0$\ are\textbf{\ }%
\begin{equation}
\lim_{_{\Lambda =0}^{r\rightarrow \infty }}R_{\theta \varphi \theta }^{~\ \
\ \ \varphi }=-3Bm~,~\lim_{_{\Lambda =0}^{r\rightarrow \infty }}R_{\theta
\varphi \varphi }^{~\ \ \ \ \theta }=3Bm\sin ^{2}\theta ~,
\end{equation}%
hence a nonvanishing parameter $B$ obstructs asymptotic flatness.

Equation (\ref{Hspec2b}) gives $\bar{G}_{2X}=0$ (unless $B=0\,$, a case
already discussed). Then Eq. (\ref{Hspec2a}) yields%
\begin{eqnarray}
\frac{\bar{G}_{2}}{2A} &=&-\frac{3mB^{2}}{r\left( 1+Br\right) }-3\left(
1+2Br\right) \Lambda  \notag \\
&&-\frac{B^{2}\left[ 11+72mB+12\left( 1+6mB\right) Br\right] }{2\left(
1+Br\right) }  \notag \\
&&-3B^{2}\left( 1\!+\!6mB\right) \left( 1\!+\!2Br\right) \ln \left\vert 
\frac{Br}{1\!+\!Br}\right\vert ~,
\end{eqnarray}%
or in terms of the scalar, 
\begin{eqnarray}
\bar{G}_{2} &=&-\frac{6mA^{2}B^{3}}{\phi -A}-6\left( 2\phi -A\right) \Lambda
\notag \\
&&+AB^{2}\left( \frac{A}{\phi }-12\right) \left( 1+6mB\right)  \notag \\
&&-6B^{2}\left( 1\!+\!6mB\right) \left( 2\phi \!-\!A\right) \ln \left\vert 
\frac{\phi \!-\!A}{\phi }\right\vert ~,
\end{eqnarray}%
and Eq. (\ref{Hspec4a}) is also verified.

\subsubsection{Horizons and singularities}

\begin{figure}[th]
\includegraphics[height=6.2cm,angle=0]{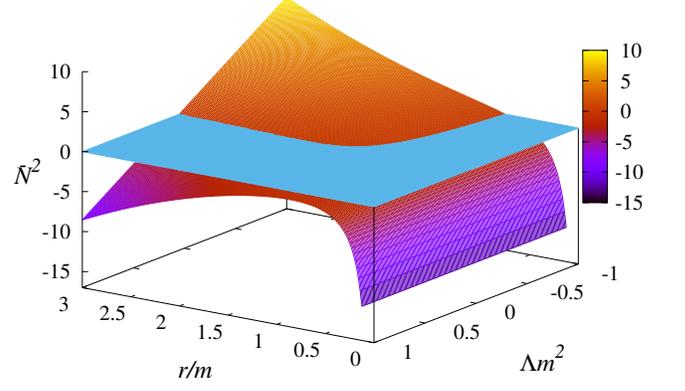} 
\caption{The metric function $\bar{N}^{2}$ represented as function of the
radial distance from the central singularity (in units of mass) and of the
parameter $\Lambda m^{2}$, for $Bm=1$. For negative values of $\Lambda $ the
singularity is hidden by a horizon (represented by the intersection of $\bar{%
N}^{2}$ with the zero plane, depicted in sky blue). For positive values of $%
\Lambda $ the metric function stays negative, representing a homogeneous
spacetime with central naked singularity.}
\label{Fig1}
\end{figure}

Although the leading terms of (\ref{N2new}) at low \thinspace $r/m$ values
may be negative, for negative $\Lambda $ and positive $B$ the cosmological
and logarithmical terms are positive, dominating the behavior of the metric
at larger distances. To illustrate this we plot the metric function with $%
Bm=1$ on Fig. \ref{Fig1}. In the parameter range with negative $\Lambda $
this represents a new spacetime with one horizon. For positive values of $%
\Lambda $ there is no horizon, $\bar{N}^{2}<0$ and a Kantowski-Sachs type
geometry emerges. 

\begin{figure}[th]
\includegraphics[height=6.2cm,angle=0]{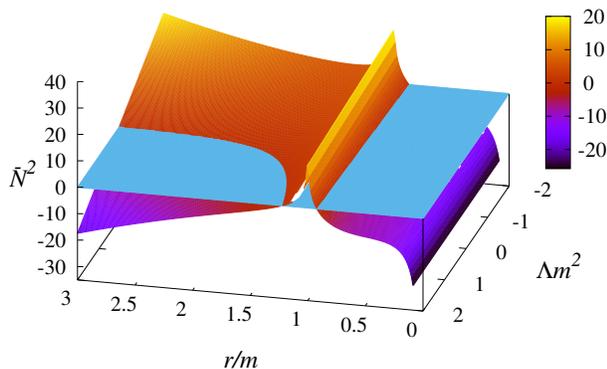} 
\caption{The metric function $\bar{N}^{2}$ represented as function of the
radial distance from the central singularity (in units of mass) and of the
parameter $\Lambda m^{2}$, for $Bm=-1$. For all values of $\Lambda $ the
central singularity is hidden by a horizon (represented by the intersection
of $\bar{N}^{2}$ with the zero plane, depicted in sky blue). Another,
logarithmic singularity is generated outside the horizon, rendering the
spacetime to a naked singularity. Outside the logarithmic singularity the
spacetime is spherically symmetric for negative $\Lambda m^{2}$ values,
while after stepping into the positive regime, another horizon appears, the
spacetime becoming homogeneous outside it.}
\label{Fig2}
\end{figure}

A similar plot for $Bm=-1$ is represented on Fig. \ref{Fig2}. There a
horizon emerges for all values of $\Lambda $, nevertheless an intriguing
feature shows up. Outside the horizon there is a singularity generated by
the blowing up of the logarithmic term in Eq. (\ref{N2new}). Outside the
logarithmic singularity the spacetime is spherically symmetric for negative $%
\Lambda m^{2}$ values, while in the positive regime another horizon appears,
rendering the spacetime homogeneous outside it. 

\begin{figure}[th]
\includegraphics[height=6.2cm,angle=0]{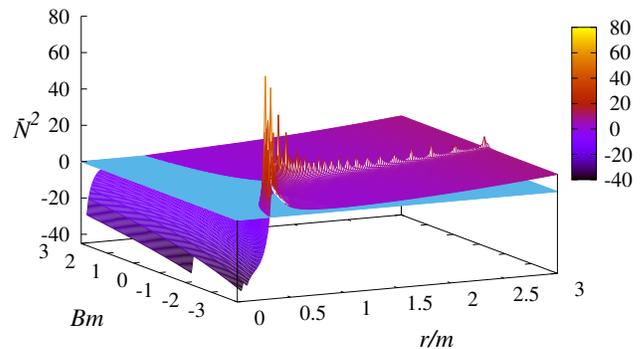} 
\caption{The metric function $\bar{N}^{2}$ represented as function of the
radial distance from the central singularity (in units of mass) and of the
parameter $Bm$, for $\Lambda m^{2}=-1$. For all values of $B$ the central
singularity is hidden by a horizon (represented by the intersection of $\bar{%
N}^{2}$ with the zero plane, depicted in sky blue). For negative values of $%
B $ however a logarithmic singularity appears outside the horizon along the
hyperbola $Br=-1$. This singularity is positive for $Bm<-1/6$ (depicted),
and negative for $-1/6<Bm<0$ (consequently for $r>6$). }
\label{Fig3}
\end{figure}

The metric coefficient for $\Lambda m^{2}=-1$ is represented as function of
the radial distance and the parameter $Bm$ on Fig. \ref{Fig3}. For positive $%
Bm$ values the black hole structure emerges again. For negative values of $%
Bm $ the logarithmic singularity (positive for $Bm<-1/6$, negative
otherwise) appears outside the horizon, at $Br=-1$. This feature has already
been encountered on Fig. \ref{Fig2}. 

\begin{figure}[th]
\includegraphics[height=6cm,angle=0]{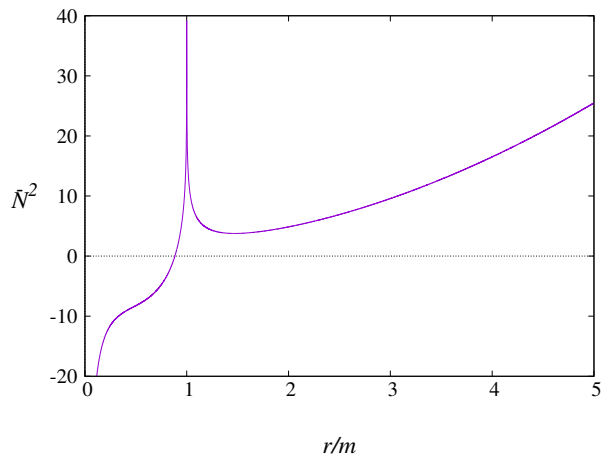} 
\caption{The metric function $\bar{N}^{2}$ represented as function of the
radial distance from the central singularity (in units of mass), for $Bm=-1$
and $\Lambda m^{2}=-1$. The logarithmic singularity appears outside the
horizon, which hides only the central singularity.}
\label{Fig4}
\end{figure}

We illustrate this singularity for $Bm=-1$ and $\Lambda m^{2}=-1$ on Fig. %
\ref{Fig4}. We checked that both the Ricci curvature scalar and the
Kretschmann scalar diverge at the logarithmic singularity, confirming the
singularity is not a coordinate artifact. 

\section{The Einstein frame description of the solutions}

In the previous section we have identified black hole, naked singularity and
homogeneous solutions for the simple Horndeski-type action

\begin{equation}
S^{EFT}=\int d^{4}x\sqrt{-\tilde{g}}\left[ G_{2}\left( \phi ,X\right)
+G_{4}\left( \phi \right) \tilde{R}\right] ~,
\end{equation}%
with the ansatz (\ref{onefunc}) of only one independent metric function. For
nonconstant $G_{4}$ this action is in Jordan frame, the natural frame of the
Horndeski-type theories. The advantage of the Jordan frame is that only the
metric couples to matter (minimally), however the coupling of the metric and
scalar is intricate.

By a conformal transformation $\widehat{g}_{ab}=\Omega ^{2}\tilde{g}_{ab}$
the expression $\sqrt{-\tilde{g}}\tilde{R}$ generates $\Omega ^{-2}\sqrt{-%
\widehat{g}}\widehat{R}$ as the only curvature term \cite{Wald}. In order to
ensure a minimal coupling of the scalar to the metric tensor, hence achieve
the Einstein frame, the conformal factor $\Omega ^{2}=\bar{G}_{4}\left( \phi
\right) >0$ should be chosen. The line element conformal to (\ref{ds2h}) in
the Einstein frame becomes%
\begin{equation}
\widehat{ds^{2}}=-\widehat{N^{2}}dt^{2}+\widehat{M^{2}}d\widehat{r}^{2}+%
\widehat{r}^{2}\left( d\theta ^{2}+\sin ^{2}\theta d\varphi ^{2}\right) ~,
\end{equation}%
with the new curvature coordinate $\widehat{r}=\bar{G}_{4}^{1/2}r$ and new
metric functions%
\begin{equation}
\widehat{N^{2}}=\bar{G}_{4}\bar{N}^{2}~,\quad \widehat{M^{2}}=\frac{\bar{M}%
^{2}}{\left[ 1+\frac{r}{2}\left( \ln \bar{G}_{4}\right) ^{\prime }\right]
^{2}}~.
\end{equation}%
In deriving these we have employed the identity%
\begin{equation}
\frac{1}{1+\frac{r}{2}\left( \ln \bar{G}_{4}\right) ^{\prime }}=1-\frac{%
\widehat{r}}{2}\frac{d\ln \bar{G}_{4}}{d\widehat{r}}~.
\end{equation}%
For the special cases discussed in Sec. \ref{new} the metric becomes more
complicated in the Einstein frame, as shown in Table \ref{Einstein}. 
\begin{table}[tbph]
\begin{center}
\begin{tabular}{c||c|c|c}
$\Omega ^{2}=\bar{G}_{4}=\phi $ & Eq. for $\bar{N}^{2}$ & $\widehat{N^{2}}$
& $\widehat{M^{2}}$ \\ \hline\hline
$\left( 16\pi G\right) ^{-1}$ & (\ref{SchdS}) & $\frac{1}{16\pi G}\bar{N}%
^{2} $ & $\bar{N}^{-2}$ \\ 
$r^{\alpha }$ & (\ref{Nalpha}) & $r^{\alpha }\bar{N}^{2}$ & $\frac{4}{\left(
2+\alpha \right) ^{2}}\bar{N}^{-2}$ \\ 
$A\left( 1+Br\right) $ & (\ref{N2new}) & $A\left( 1+Br\right) \bar{N}^{2}$ & 
$\frac{4\left( 1+Br\right) ^{2}}{\left( 2+3Br\right) ^{2}}\bar{N}^{-2}$%
\end{tabular}%
\end{center}
\caption{The metric in Einstein frame in the particular cases investigated,
the Einstein frame and Jordan frame curvature coordinates being related as $%
\widehat{r}=rG_{4}^{1/2}$. }
\label{Einstein}
\end{table}

\section{Concluding remarks}

In this paper we have explored the recently developed 2+1+1 decomposition of
spacetime \cite{GKG}, based on a nonorthogonal double foliation, for the
study of spherically symmetric, static solutions of a particular subclass of
the Horndeski scalar-tensor theory. This subclass contains k-essence models
nonminimally coupled to the metric (however the coupling depends only on the
scalar field, not its derivatives).

We started the analysis by employing the approach of the effective field
theory (EFT) of modified gravity, in which the action is conceived as a
functional of a sufficiently large set of scalars, constructed from metric
and embedding variables, all adapted to the nonorthogonal 2+1+1
decomposition. The choice of variables roughly followed and generalized the
earlier analysis \cite{KGT}, performed in the orthogonal double foliation
particular case, with the notable inclusion of a nonorthogonality parameter.

First, we proved that the class of Horndeski Lagrangians with $G_{5}=0$ (the
inclusion of the latter would be incompatible with observations) can be
expressed in this EFT form (in terms of the scalars constructed from the
metric and embedding variables adapted to the nonorthogonal double
foliation).

Next, by studying the first order perturbation of the EFT action, we derived
three equations of motion for the metric and embedding variables, which
reduce to those derived earlier in an orthogonal 2+1+1 decomposition, and a
fourth equation for the metric parameter $\mathcal{N}$ related to the
nonorthogonality of the foliation. An additional equation emerged for the
scalar field. As a check, we recovered the Schwarzschild solution for the
Einstein--Hilbert action rewritten in this framework.

Then, for the Horndeski class of theories with vanishing $G_{3}$ and $G_{5}$%
, but generic functions $G_{2}\left( \phi ,X\right) $ (k-essence) and $%
G_{4}\left( \phi \right) $ (scalar field dependent nonminimal coupling to
the metric) we proved the unicity theorem, according to which no action
beyond Einstein--Hilbert allows for the Schwarzschild solution. With this,
for a spherical symmetric and static setup, we extended the unicity theorem
previously announced in the literature, which applies for a function $G_{2}$
linear in $X$ (asymptotycally homogeneous in $X$) and asymptotical flatness.

After that, we integrated the EFT field equations for the case with only one
independent metric function. We assumed $G_{4}\left( \phi \right) =\phi $,
while $G_{2}\left( \phi ,X\right) $ was fixed by the equations of motion. We
discussed in particular scalar fields with polynomial and linear radial
dependences, obtaining new solutions characterized by mass, tidal charge, a
parameter generalizing both, the cosmological constant and an additional
parameter. We have also written them up as conformally related metrics in
the Einstein frame.

These solutions represent naked singularities, black holes or have the
double horizon structure of the Schwarzschild--de Sitter spacetime.
Solutions with homogeneous Kantowski--Sachs type regions also emerged.
Finally, one of the solutions obtained for the function $G_{4}$ linear in
the curvature coordinate in certain parameter range exhibits an intriguing
logarithmic singularity lying outside the horizon. All these solutions were
asymptotically nonflat (even when the cosmological constant was switched
off). Hence, the black hole solutions evade previously known unicity
theorems, and exhibit scalar hair.

\section*{Acknowledgements}

This work was supported by the Hungarian National Research Development and
Innovation Office (NKFIH) in the form of the Grant No. 123996 and has been
carried out in the framework of COST actions CA16104 (GWverse) and CA18108
(QG-MM), supported by COST (European Cooperation in Science and Technology).
C.N. was supported by the UNKP-19-3\textbf{\ }and UNKP-20-3 New National
Excellence Programs of the Ministry for Innovation and Technology. Z.K. was
supported by the J\'{a}nos Bolyai Research Scholarship of the Hungarian
Academy of Sciences; also by the UNKP-19-4 and UNKP-20-5 New National
Excellence Programs of the Ministry for Innovation and Technology.

\ \appendix

\section{The relation among the scalar variables in the two bases\label{App0}%
}

From Table III. and Eqs. (34) of Ref. \cite{GKG}, the tensorial and
vectorial embedding variables turn out to be related as%
\begin{eqnarray}
K_{ab}^{\ast } &=&\frac{1}{\mathfrak{c}}\left( K_{ab}+\mathfrak{s}%
L_{ab}\right) ,  \notag \\
L_{ab}^{\ast } &=&\frac{1}{\mathfrak{c}}\left( L_{ab}-\mathfrak{s}%
K_{ab}\right) ~,  \notag \\
\mathcal{K}_{a}^{\ast } &=&\mathcal{K}_{a}+\frac{\mathfrak{s}}{\mathfrak{c}}%
D_{a}\ln \frac{N}{\mathfrak{c}M}~,  \notag \\
\mathcal{L}_{a}^{\ast } &=&\mathcal{L}_{a}+\frac{\mathfrak{s}}{\mathfrak{c}}%
D_{a}\ln \frac{N}{\mathfrak{c}M}~,
\end{eqnarray}%
from which the relations connecting the sets of scalars (\ref{scalars1}) and
(\ref{scalars2}) emerge: 
\begin{eqnarray}
\varkappa ^{\ast } &=&\frac{1}{\mathfrak{c}^{2}}\left( \varkappa +2\mathfrak{%
s}K_{ab}L^{ab}+\mathfrak{s}^{2}\lambda \right) ~,  \notag \\
\lambda ^{\ast } &=&\frac{1}{\mathfrak{c}^{2}}\left( \lambda -2\mathfrak{s}%
K_{ab}L^{ab}+\mathfrak{s}^{2}\varkappa \right) ~,  \notag \\
\mathfrak{K}^{\ast } &=&\mathfrak{K}+2\frac{\mathfrak{s}}{\mathfrak{c}}%
\mathcal{K}^{a}D_{a}\ln \frac{N}{\mathfrak{c}M}+\left( \frac{\mathfrak{s}}{%
\mathfrak{c}}\right) ^{2}\left( D_{a}\ln \frac{N}{\mathfrak{c}M}\right)
^{2}~,  \notag \\
\mathfrak{k}^{\ast } &=&\mathfrak{k}+2\frac{\mathfrak{s}}{\mathfrak{c}}%
\mathcal{L}^{a}D_{a}\ln \frac{N}{\mathfrak{c}M}+\left( \frac{\mathfrak{s}}{%
\mathfrak{c}}\right) ^{2}\left( D_{a}\ln \frac{N}{\mathfrak{c}M}\right)
^{2}~,  \notag \\
K^{\ast } &=&\frac{1}{\mathfrak{c}}\left( K+\mathfrak{s}L\right) ~,  \notag
\\
L^{\ast } &=&\frac{1}{\mathfrak{c}}\left( L-\mathfrak{s}K\right) ~,
\label{1}
\end{eqnarray}%
where $\left( D_{a}F\right) ^{2}\equiv \left( D_{a}F\right) \left(
D^{a}F\right) $ for any function $F$.

Similarly, the scalars (\ref{KLscalar}) and (\ref{KLstarscalar}) are related
through

\begin{eqnarray}
\mathcal{K}^{\ast } &=&\frac{1}{\mathfrak{c}}\left( \mathcal{K}-\mathfrak{s}%
\mathcal{L}\right) +\frac{\mathfrak{c}}{M}\left( \partial _{\chi
}-M^{a}D_{a}\right) \left( \frac{\mathcal{N}}{N}\right) ~,  \notag \\
\mathcal{L}^{\ast } &=&\frac{1}{\mathfrak{c}}\left( \mathfrak{s}\mathcal{K}+%
\mathcal{L}\right) +\frac{\mathfrak{c}^{2}}{N}\left( \partial
_{t}-N^{a}D_{a}\right) \left( \frac{\mathcal{N}}{N}\right) ~~.  \label{2}
\end{eqnarray}%
These emerge from Table III. of Ref. \cite{GKG} and the inverse relations%
\begin{eqnarray}
n^{a} &=&\frac{1}{N}\left( \frac{\partial }{\partial t}\right) ^{a}-\frac{%
\mathfrak{s}}{\mathfrak{c}}\frac{1}{M}\left( \frac{\partial }{\partial \chi }%
\right) ^{a}-\frac{1}{N}N^{a}+\frac{\mathfrak{s}}{\mathfrak{c}}\frac{1}{M}%
M^{a}~,  \notag \\
m^{a} &=&\frac{1}{M}\left( \frac{\partial }{\partial \chi }\right) ^{a}-%
\frac{1}{M}M^{a}~,
\end{eqnarray}%
and%
\begin{eqnarray}
l^{a} &=&\frac{\mathfrak{s}}{N}\left( \frac{\partial }{\partial t}\right)
^{a}+\frac{1}{\mathfrak{c}M}\left( \frac{\partial }{\partial \chi }\right)
^{a}-\frac{\mathfrak{s}}{N}N^{a}-\frac{1}{\mathfrak{c}M}M^{a}~,  \notag \\
k^{a} &=&\frac{\mathfrak{c}}{N}\left( \frac{\partial }{\partial t}\right)
^{a}-\frac{\mathfrak{c}}{N}N^{a}~
\end{eqnarray}%
of Eqs. (\ref{ddt})-(\ref{ddchig}), giving 
\begin{eqnarray}
l^{a}-\mathfrak{s}n^{a} &=&\frac{\mathfrak{c}}{M}\left[ \left( \frac{%
\partial }{\partial \chi }\right) ^{a}-M^{a}\right] ~,  \notag \\
\mathfrak{s}l^{a}+n^{a} &=&\frac{\mathfrak{c}^{2}}{N}\left[ \left( \frac{%
\partial }{\partial t}\right) ^{a}-N^{a}\right] ~.
\end{eqnarray}

Finally, from the last equality of Eq. (\ref{KLform}) one obtains 
\begin{equation}
\mathfrak{K}=\mathfrak{k}-2\mathfrak{c}^{2}\mathcal{L}^{a}D_{a}\left( \frac{%
\mathcal{N}}{N}\right) +\mathfrak{c}^{4}\left[ D_{a}\left( \frac{\mathcal{N}%
}{N}\right) \right] ^{2}.  \label{3}
\end{equation}

Remarkably, in the particular subcase of orthogonal foliations ($\mathfrak{s}%
=0=\mathcal{N}$ and $\mathfrak{c}=1$) all starry scalars coincide with their
unstarred versions in Eqs. (\ref{1}) and (\ref{2}), while Eq. (\ref{3})
implies a further simplification $\mathfrak{K}=\mathfrak{k}$, further
reducing the number of independent scalars.

For nonorthogonal foliations the number of independent embedding scalars is
reduced to 7 by Eqs. (\ref{1})-(\ref{3}), which also involve the three
metric components $N$, $M$, and $\mathcal{N}$.

\section{The decomposed Horndeski Lagrangian in the $\left( n,m\right) ~$%
basis\label{App1}}

We give in this appendix the expressions of $\tilde{\square}\phi $ and $%
L_{4}^{\mathrm{H}^{\prime }}$ in terms of the kinematical quantities arising
in the $\left( n,m\right) ~$basis. For this first we calculate%
\begin{eqnarray}
\tilde{\nabla}_{a}\tilde{\nabla}_{b}\phi &=&\mathfrak{s}\sqrt{X}\left(
K_{ab}+2m_{(a}\mathcal{K}_{b)}+m_{a}m_{b}\mathcal{K}-n_{a}\mathfrak{a}%
_{b}\right.  \notag \\
&&\left. +n_{a}m_{b}\mathcal{L}^{\ast }\right) +\mathfrak{c}\sqrt{X}\left(
L_{ab}^{\ast }+n_{a}\mathcal{L}_{b}^{\ast }+n_{b}\mathcal{K}_{a}\right. 
\notag \\
&&\left. +n_{a}n_{b}\mathcal{L}^{\ast }+m_{a}\mathfrak{b}_{b}^{\ast
}+m_{a}n_{b}\mathcal{K}\right) +\sqrt{X}n_{b}\tilde{\nabla}_{a}\mathfrak{s} 
\notag \\
&&+\sqrt{X}m_{b}\tilde{\nabla}_{a}\mathfrak{c}+\frac{1}{2X}\left( \mathfrak{s%
}n_{a}+\mathfrak{c}m_{a}\right) \left( \mathfrak{s}n_{b}+\mathfrak{c}%
m_{b}\right)  \notag \\
&&\times \tilde{\nabla}^{c}\phi \tilde{\nabla}_{c}X+\sqrt{X}\left( \mathfrak{%
s}n_{b}+\mathfrak{c}m_{b}\right) \left( \mathfrak{s}n^{c}+\mathfrak{c}%
m^{c}\right)  \notag \\
&&\times \left( n_{a}\tilde{\nabla}_{c}\mathfrak{s+}m_{a}\tilde{\nabla}_{c}%
\mathfrak{c}\right) +\sqrt{X}\left( \mathfrak{c}n_{a}+\mathfrak{s}%
m_{a}\right)  \notag \\
&&\times \left( \mathfrak{s}n_{b}+\mathfrak{c}m_{b}\right) \left( \mathfrak{c%
}\mathcal{K}-\mathfrak{s}\mathcal{L}^{\ast }\right) +\sqrt{X}\left( 
\mathfrak{s}n_{b}+\mathfrak{c}m_{b}\right)  \notag \\
&&\times \left[ \mathfrak{sc}\left( \mathcal{K}_{a}-\mathcal{L}_{a}^{\ast
}\right) +\mathfrak{s}^{2}D_{a}\left( \ln N\right) \right.  \notag \\
&&\left. -\mathfrak{c}^{2}D_{a}\left( \ln M\right) \right] ~.
\end{eqnarray}%
Hence,%
\begin{eqnarray}
\tilde{\square}\phi &=&\frac{\tilde{\nabla}^{a}\phi \tilde{\nabla}_{a}X}{2X}+%
\sqrt{X}\left[ \mathfrak{s}\left( K+\mathcal{K}\right) \right.  \notag \\
&&\left. +\mathfrak{c}\left( L^{\ast }-\mathcal{L}^{\ast }\right) \mathfrak{%
+c}^{2}\left( \mathfrak{c}n^{a}+\mathfrak{s}m^{a}\right) \tilde{\nabla}_{a}%
\frac{\mathfrak{s}}{\mathfrak{c}}\right] ~
\end{eqnarray}%
and%
\begin{eqnarray}
L_{4}^{\mathrm{H}^{\prime }} &=&G_{4}\left[ R+K_{ab}K^{ab}-L_{ab}^{\ast
}L^{\ast ab}-K^{2}+\left( L^{\ast }\right) ^{2}\right]  \notag \\
&&+2\sqrt{X}G_{4\phi }\left[ \mathfrak{s}\left( K+\mathcal{K}\right) +%
\mathfrak{c}\left( L^{\ast }-\mathcal{L}^{\ast }\right) \right.  \notag \\
&&\left. +\mathfrak{c}^{2}\left( \mathfrak{c}n^{a}+\mathfrak{s}m^{a}\right) 
\tilde{\nabla}_{a}\frac{\mathfrak{s}}{\mathfrak{c}}\right]  \notag \\
&&-\left( G_{4}-2XG_{4X}\right) \left[ 2K\mathcal{K}+2L^{\ast }\mathcal{L}%
^{\ast }\right.  \notag \\
&&-2\left( \mathcal{K}^{a}+\mathfrak{c}^{2}D^{a}\frac{\mathfrak{s}}{%
\mathfrak{c}}\right) \left( \mathcal{K}_{a}+\mathfrak{c}^{2}D_{a}\frac{%
\mathfrak{s}}{\mathfrak{c}}\right)  \notag \\
&&+2\mathfrak{c}^{2}\left( \mathfrak{c}K+\mathfrak{s}L^{\ast }\right) \left( 
\mathfrak{s}n^{a}+\mathfrak{c}m^{a}\right) \tilde{\nabla}_{a}\frac{\mathfrak{%
s}}{\mathfrak{c}}  \notag \\
&&-2\mathfrak{c}^{2}\left( \mathfrak{c}L^{\ast }\mathfrak{+s}K\right) \left( 
\mathfrak{c}n^{a}+\mathfrak{s}m^{a}\right) \tilde{\nabla}_{a}\frac{\mathfrak{%
s}}{\mathfrak{c}}  \notag \\
&&\left. -D^{b}\left( \ln \frac{N}{\mathfrak{c}}\right) D_{b}\ln \left( 
\mathfrak{c}M\right) \right] ~  \notag \\
&&+2XG_{4X}\left[ \mathfrak{s}^{2}K_{ab}K^{ab}+\mathfrak{c}^{2}L_{ab}^{\ast
}L^{\ast ab}-\mathfrak{s}^{2}K^{2}\right.  \notag \\
&&\left. -\mathfrak{c}^{2}\left( L^{\ast }\right) ^{2}+2\mathfrak{sc}%
L_{ab}^{\ast }K^{ab}-2\mathfrak{sc}L^{\ast }K\right] ~.
\end{eqnarray}

These equations, when compared with Eqs. (\ref{DAlembertphi}) and (\ref%
{Hornkl}), clearly show that the Hordenski action takes simpler form in
terms of the kinematical quantities defined in the $\left( k,l\right) $
basis. This is because the scalar function $\phi $ depends only on $\chi $;
therefore it is adapted to $l^{a}$.

\section{The field equations in the $\left( n,m\right) ~$basis\label{App2}}

By choosing the $\left( n,m\right) $ basis as primary, the action is
naturally rewritten as depending on the variables 
\begin{equation}
S^{EFT(2)}\left[ \mathcal{N},N,M,\mathcal{K},\mathfrak{K},K,\varkappa ,%
\mathcal{L}^{\ast },L^{\ast },\lambda ^{\ast },R,\phi \right] ~.
\label{EFTalt}
\end{equation}%
This action depends on the scalar $\phi $, the metric variables $\left( 
\mathcal{N},N,M\right) $, the induced curvature scalar$\ R$ and the
embedding scalars $\left( \mathcal{K},\mathfrak{K},K,\varkappa ,\mathcal{L}%
^{\ast },L^{\ast },\lambda ^{\ast }\right) $ related to the $\left(
n,m\right) $ basis. Beside switching the starry / nonstarry variables when
going from the $\left( k,l\right) $ basis to the $\left( n,m\right) $ basis
the only step needing additional explanation is the change of the variable $%
\mathfrak{k}$ into $\mathfrak{K}$, as the complementary variables $\mathfrak{%
k}^{\ast }$ into $\mathfrak{K}^{\ast }$ are not formed from normal
fundamental forms. Note that cf. Appendix \ref{App0} the set $\left( 
\mathcal{K},\mathfrak{K},K,\varkappa ,\mathcal{L}^{\ast },L^{\ast },\lambda
^{\ast }\right) $ cannot be transformed to the alternative set $\left( 
\mathcal{K}^{\ast },\mathfrak{k},K^{\ast },\varkappa ^{\ast },\mathcal{L}%
,L,\lambda \right) $ employed in the main part of the paper without the use
of the new variables $K_{ab}L^{ab}$ and $\mathcal{L}^{a}D_{a}\left( \frac{%
\mathcal{N}}{N}\right) $ appearing in the transformations (\ref{1}) and (\ref%
{3}), respectively.

A similar variation procedure of the EFT action (\ref{EFTalt}) as described
in Sec. \ref{firstO} yields the following field equations:%
\begin{gather}
L_{\mathcal{N}}^{EFT(2)}+\,\frac{\partial _{r}\left( r^{2}L_{\mathcal{K}%
}^{EFT(2)}\right) }{r^{2}\bar{N}\bar{M}}-\,\frac{2L_{K}^{EFT(2)}}{r\bar{N}%
\bar{M}}=0~,  \notag \\
\bar{L}^{EFT(2)}+\bar{N}L_{N}^{EFT(2)}+\frac{\mathbf{D}_{r}L_{\mathcal{L}%
^{\ast }}^{EFT(2)}}{\bar{M}}=0~,  \notag \\
\bar{L}^{EFT(2)}+\bar{M}L_{M}^{EFT(2)}-\frac{2\mathcal{F}^{\left( 2\right) }%
}{r\bar{M}}+\frac{\partial _{r}\bar{N}}{\bar{M}\bar{N}}L_{\mathcal{L}^{\ast
}}^{EFT(2)}=0~,  \notag \\
\bar{L}^{EFT(2)}-\frac{\mathbf{D}_{r}\mathcal{F}^{\left( 2\right) }}{{\bar{M}%
}}-\frac{2L_{R}^{EFT(2)}}{r^{2}}=0~,  \label{EOMalt}
\end{gather}%
where%
\begin{eqnarray}
\mathbf{D}_{r} &=&\frac{2}{r}+\frac{\partial _{r}\bar{N}}{\bar{N}}+\partial
_{r}~,  \notag \\
\mathcal{F}^{\left( 2\right) } &=&L_{L^{\ast }}^{EFT\left( 2\right) }+\frac{2%
}{\bar{M}r}L_{\lambda ^{\ast }}^{EFT\left( 2\right) }~.
\end{eqnarray}%
It is not simple to show the equivalence of these to the set (\ref{EOM1})-(%
\ref{EOM4}), although the last three equations are formally quite similar in
the two approaches and both reduce to the same set in the orthogonal double
foliation limit. The comparison is nontrivial as $L^{EFT}$ and $L^{EFT\left(
2\right) }$ depend on different sets of variables (and there is need for
additional variables to relate them, as emphasized above). Beside also they
differ by total covariant divergencies (obvious from the procedure they were
derived). Taking into account these should lead to the same set of solutions.

\end{document}